\documentclass[3p,number,sort&compress]{elsarticle}

\usepackage{natbib} 
\usepackage{geometry}
\usepackage{graphicx}
\usepackage{hyperref}

\usepackage{amsmath,amssymb} 
\usepackage{siunitx}
\usepackage{bm}
\usepackage{url}
\usepackage[version=3]{mhchem} 
\usepackage{subcaption}
\usepackage{color}
\usepackage{comment}

\newcommand{\figref}[1]{Fig.\,\ref{#1}}
\newcommand{\figrefs}[2]{Figs.\,\ref{#1} and \ref{#2}}
\newcommand{\Figref}[1]{Figure\,\ref{#1}}
\newcommand{\Figrefs}[2]{Figures\,\ref{#1} and \ref{#2}}
\newcommand{\tabref}[1]{Tab.\,\ref{#1}}
\newcommand{\secref}[1]{Sec.\,\ref{#1}}
\newcommand{\eref}[1]{Eq.\,(\ref{#1})}
\newcommand{\dd}{\mathrm{d}}

\let\today\relax
\makeatletter
\def\ps@pprintTitle{%
    \let\@oddhead\@empty
    \let\@evenhead\@empty
    \def\@oddfoot{\footnotesize\itshape
         { } \hfill\today}%
    \let\@evenfoot\@oddfoot
    }
\makeatother

\begin{document}

\title{Simultaneous Modeling of In Vivo and In Vitro Effects of  Nondepolarizing Neuromuscular Blocking Drugs}

\author[1]{Hikaru Hoshino\corref{cor1}} 
\ead{hoshino@eng.u-hyogo.ac.jp}

\author[1,2]{Eiko Furutani}

\cortext[cor1]{Corresponding author}

\address[1]{Department of Electrical Materials and Engineering, University of Hyogo, Hyogo, Japan}
\address[2]{Department of Anesthesiology, Kagawa University, Kagawa, Japan }

\begin{abstract}
Nondepolarizing neuromuscular blocking drugs (NDNBs) are clinically used to produce muscle relaxation during general anesthesia.
This paper explores a suitable model structure to simultaneously describe \emph{in vivo} and \emph{in vitro} effects of three clinically used NDNBs, cisatracurium, vecuronium, and rocuronium.  
In particular, it is discussed how to reconcile an apparent discrepancy that rocuronium is less potent at inducing muscle relaxation \emph{in vivo} than predicted from \emph{in vitro} experiments. 
We develop a framework for estimating model parameters from published \emph{in vivo} and \emph{in vitro} data, and thereby compare the descriptive abilities of several candidate models.
{\color{red}
It is found that modeling of dynamic effect of activation of acetylcholine receptors (AChRs) is essential for describing \emph{in vivo} experimental results, and a cyclic gating scheme of AChRs is suggested to be appropriate. 
Furthermore, it is shown that the above discrepancy in experimental results can be resolved when we consider the fact that} the \emph{in vivo} concentration of ACh is quite low to activate only a part of AChRs, whereas more than $\SI{95}{\%}$ of AChRs are activated during \emph{in vitro} experiments, and that the site-selectivity is smaller for rocuronium than those for cisatracurium and vecuronium. 
\end{abstract}


\maketitle

\section{Introduction} \label{sec:intro}

Nondepolarizing neuromuscular blocking drugs (NDNBs) interrupt synaptic transmission at the neuromuscular junction and clinically used during general anesthesia to produce muscle relaxation \cite{pardo18}.  
Neuromuscular transmission is initiated by arrival of an impulse at motor nerve terminal and subsequent release of acetylcholine (ACh) molecules to the synaptic cleft. 
A part of released ACh molecules bind to nicotinic ACh receptors (AChRs) on post-junctional membranes and thereby causes a change of membrane conductance due to channel opening of AChRs, followed by occurrence  of action potential at muscle fibers and muscle contraction. 
It is well known that NDNBs act by competing with ACh for post-junctional AChRs and preventing changes in membrane conductance \cite{fagerlund09}. 
Each AChR has two non-identical binding sites, and the binding of only one molecule of NDNB is needed to prevent activation of the receptor, whereas two molecules of ACh are necessary for activation. 

While clinical effects of NDNBs are usually modeled by pharmacokinetic and pharmacodynamic (PKPD) analysis \citep{sheiner79, plaud95, fisher97,
bergeron01, kleijn11},  it is a rather black-box approach. 
To better understand clinical properties of NDNBs, several mechanism-based models have been proposed.
One of the most basic models is the two-site binding model~\citep{sine81}, which is derived based on the assumption that the effect of a drug is proportional to the fractional amount of receptors occupied by the antagonist. 
This assumption is valid in most \emph{in vitro}  experiments, and the two-site binding model has been widely used to represent  such \emph{in vitro} experimental results \cite{sine81,colquhoun88,liu08,liu09}. 
However, it is known that these \emph{in vitro} results do not directly explain clinical or \emph{in vivo} results. 
For example, although the values of $\mathrm{IC}_{50}$, the concentration needed to produce a $50 \%$ inhibition of the experimental current, take similar values for three clinically used NDNBs, cisatracurium, vecuronium, and rocuronium ($\SI{10}{nM}$, $\SI{15}{nM}$, and $\SI{17}{nM}$, respectively \cite{liu08}),  the value of $\mathrm{EC}_{50}$,  the concentration needed to produce a $50 \%$ decrease of clinically observed muscle response, for rocuronium is much higher ($\SI{1.35}{\mu M}$ \cite{plaud95}) than those for cisatracurium ($\SI{0.12}{\mu M}$ \cite{bergeron01}) and vecuronium  ($\SI{0.26}{\mu M}$ \cite{fisher97}). 
That is, rocuronium is less potent at inducing muscle relaxation \emph{in vivo} than directly predicted from \emph{in vitro} experiments.

The purpose of this paper is to develop a model describing \emph{in vivo} and \emph{in vitro} experimental results in a consistent manner by considering molecular mechanisms of the effects of NDNBs. 
Although the two-site binding model effectively describes \emph{in vitro} experimental results, it represents only static properties of an NDNBs at an equilibrium condition realized by \emph{in vitro} experimental settings.  
Since the free concentration of ACh and the degree of ACh occupancy on the receptors do not reach equilibrium during a synaptic event \citep{demazumder01},  a dynamic modeling is required to represent molecular processes of competition between ACh and NDNB molecules. 
In this direction, Dilger and coworkers \citep{wenningmann01,demazumder01,demazumder08} conducted kinetic measurements using a rapid perfusion system to determine association and dissociation rate constants of NDNB bindings.   
Particularly, a dynamic simulation was performed in \cite{demazumder01} to reproduce the time course of experimental currents by using an ordinary differential equation model.  
Furthermore, Nigrovic and Amann \cite{nigrovic03} proposed a model of neuromuscular transmission, which is termed as the \emph{competitive kinetic model} in this paper, to simulate \emph{in vivo} muscular response. 
Based on these studies, this paper addresses simultaneous modeling of \emph{in vivo} and \emph{in vitro} effects of NDNBs. 

The contribution of this paper is {\color{red}as follows. 
%
First, we propose a modification to the competitive kinetic model.} 
Specifically, we introduce a cyclic scheme for gate opening and closing of AChRs upon association and dissociation of ACh molecules: 1) agonists bind, 2) AChRs open, 3) agonists dissociate, and then 4) AChRs return to the resting condition. 
This modification is based on similar models developed for explaining the processes of desensitization of AChRs \cite{katz57,auerbach98}.
Although a reciprocal gating scheme, where agonists dissociate after AChR return to the resting condition, has been a widely accepted mechanism \cite{akk96,grosman01,auerbach12,gupta17}, the results of this paper indicates a possibility of the cyclic gating scheme. 
{\color{red}
Second, it is shown that the above apparent discrepancy in  $\mathrm{IC}_{50}$ and  $\mathrm{EC}_{50}$ can be 
explained when we consider the fact that} the \emph{in vivo} concentration of ACh is relatively low to activate only a part of AChRs, whereas more than $\SI{95}{\%}$ of AChRs are activated during \emph{in vitro} experiments, and that the site-selectivity is smaller for rocuronium than those for cisatracurium and vecuronium. 

The rest of this paper is organized as follows. 
\secref{sec:methods} introduces the two-site binding model and the competitive kinetic model with reciprocal and cyclic gating scheme and presents a framework for estimating model parameters to compare these candidate models. 
\secref{sec:results} provides  the result of parameter estimation for each model and perform an additional numerical analysis  to clarify the difference among these models. 
In \secref{sec:discussion}, we summarize the obtained results and discuss how the apparent discrepancy in  $\mathrm{IC}_{50}$ and  $\mathrm{EC}_{50}$ can be resolved, and finally \secref{sec:conclusion} concludes this paper.


\section{Methods} \label{sec:methods}

\subsection{Models of Neuromuscular Response} 

\begin{figure}[!t]
 \centering
  \includegraphics[width=0.85\linewidth]{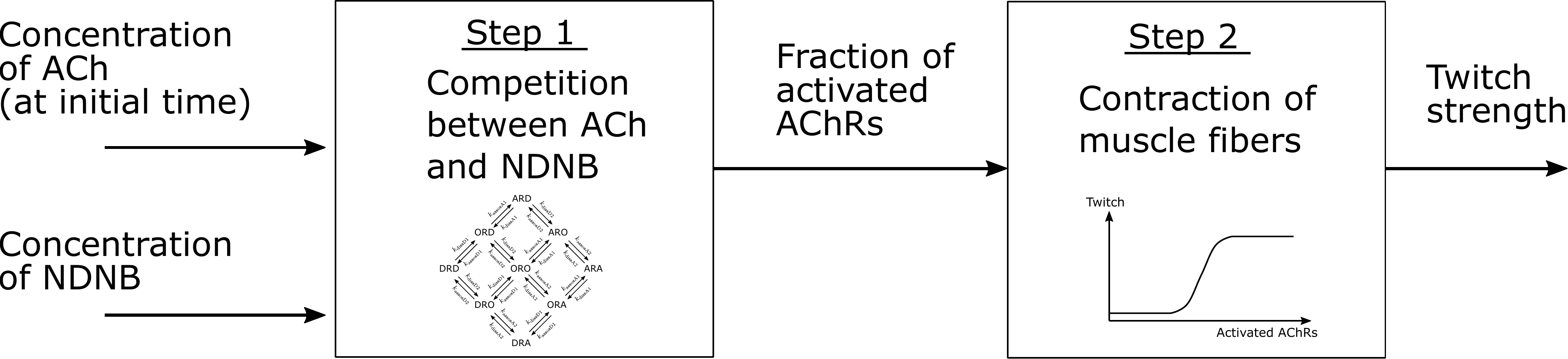}
 \caption{Overall structure of the studied models consisting of the two steps: 1) calculation of the fraction of activated AChRs after the release of ACh and 2) calculation of  the strength of the clinically observed muscle response to a stimulus. }
\label{fig:model_framework} 
\end{figure}

Here we introduce the models of neuromuscular response studied in this paper. 
The overall structure, which is based on \cite{nigrovic03} and common to all the considered models, is shown in  \figref{fig:model_framework}. 
The list of parameters are provided in \tabref{tab:parameters}. 
As shown in the figure, the \emph{in vivo} effects of NDNBs are simulated by the following two steps: 1) calculation of the fraction of activated AChRs after the release of ACh due to a stimulus to the motor nerve and 2) calculation of  the twitch strength, i.e. the strength of the clinically observed muscle response to a stimulus.

In this paper, we consider three different model structures for the first step of the above procedure: a) two-site binding model, b) competitive kinetic model with reciprocal gating scheme, and c) competitive kinetic model with cyclic gating scheme. 
Among them, the two-site binding model is a simple receptor binding model \citep{sine81}, and the concentration of activated AChRs,  $\ce{[R^\ast]}$, relative to the control value, $\ce{[R^\ast]_0}$, at the absence of NDNB is given by 
\begin{align} \label{eq:two-site_binding_model}
 \dfrac{ \ce{[R^\ast]} }{ \ce{[R^\ast]_0 } } = \dfrac{ K_\mathrm{D1} K_\mathrm{D2} }{ K_\mathrm{D1} K_\mathrm{D2} + K_\mathrm{D1} \ce{[D]} + K_\mathrm{D2} \ce{[D]} + \ce{[D]}^2 }
\end{align}
where $\ce{[D]}$ stands for the drug concentration, and $K_\mathrm{D1}$ and $K_\mathrm{D2}$ for the dissociation equilibrium constants for NDNBs binding to the first and second sites of an AChR, respectively.
Note that the right-hand side of \eref{eq:two-site_binding_model} represents the fractional amount of free AChRs not occupied by NDNB at an equilibrium condition, which can be derived based on the law of mass action.

\begin{figure}[!t]
\centering
 \begin{minipage}[t]{0.49\linewidth}
  \centering
  \includegraphics[width=0.9\linewidth]{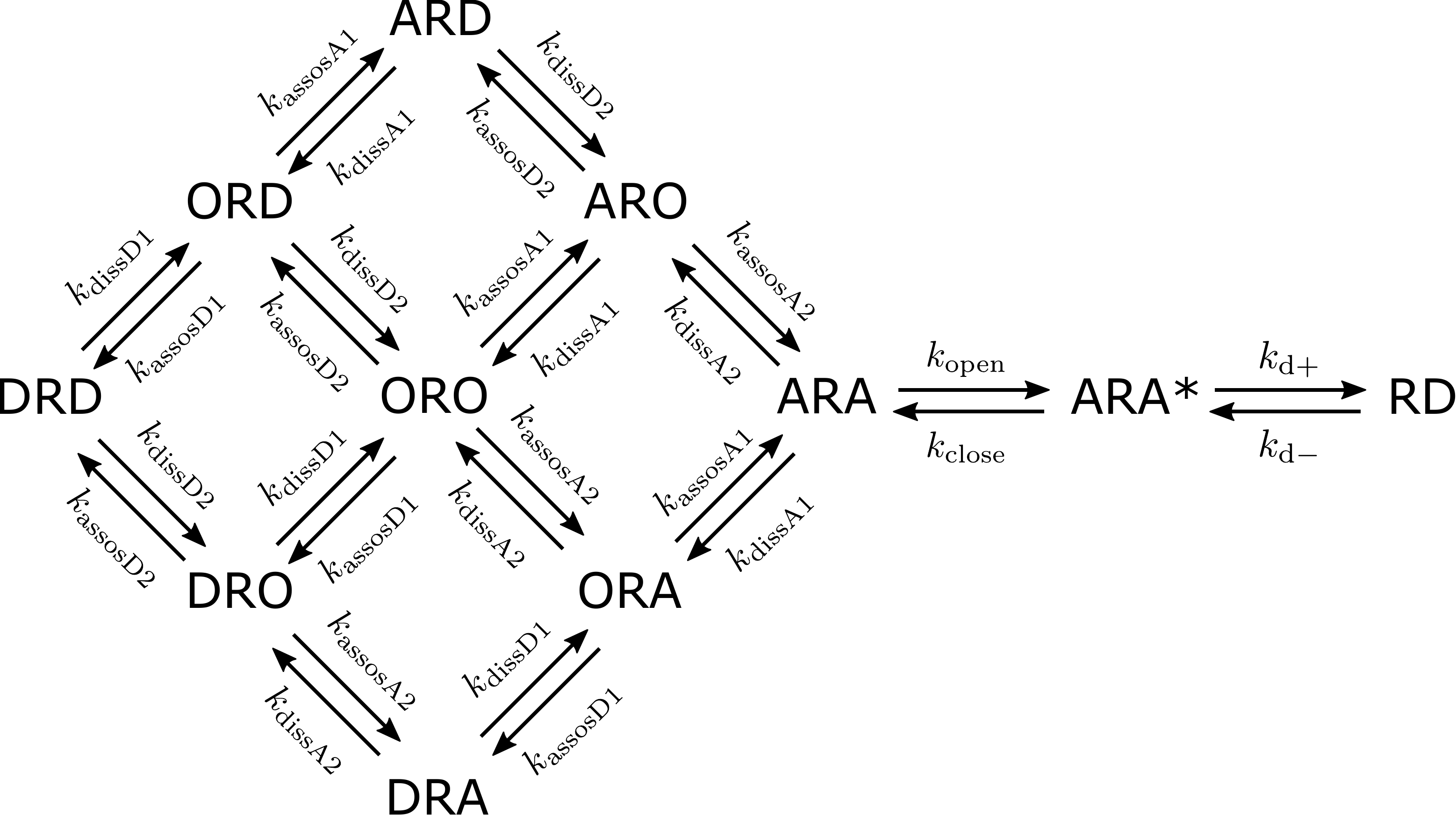}
  \subcaption{Reciprocal scheme} \label{fig:reciprocal_scheme}
 \end{minipage}
 \begin{minipage}[t]{0.49\linewidth}
  \centering
  \includegraphics[width=0.9\linewidth]{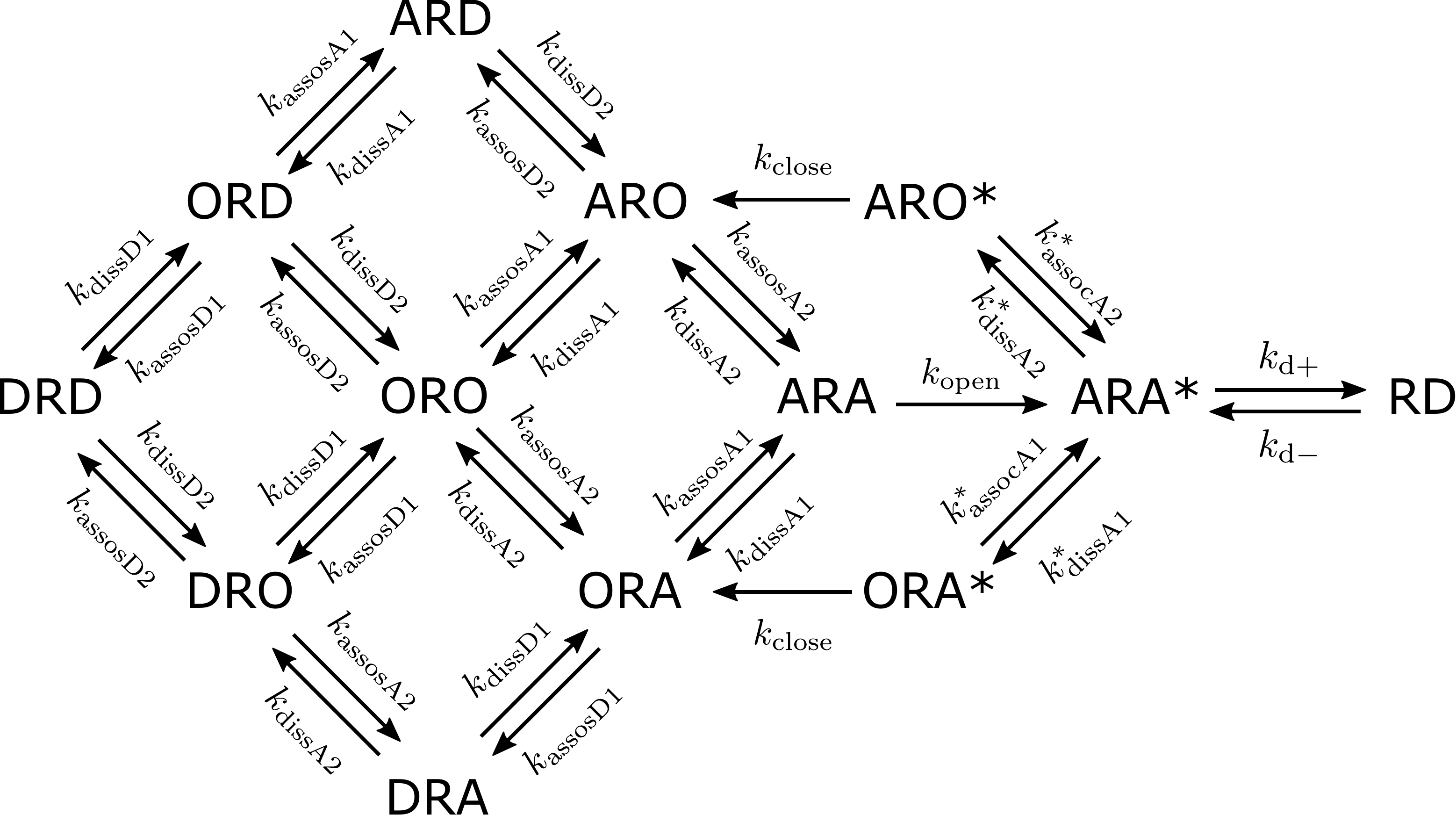}
  \subcaption{Cyclic scheme} \label{fig:cyclic_scheme}
 \end{minipage}
\caption{Two schemes for gate opening and closing of AChRs. The complexes formed by binding of ACh, denoted by $\mathrm{A}$, and NDNB, by $\mathrm{D}$, to  AChR,  by $\mathrm{R}$, are represented by 3-letter symbols, and  the symbol $\ce{ARA}^\ast$ for AChRs stands for the open state due to the conformational change of AChRs.  The symbol $\mathrm{RD}$ represents the desensitized state. 
{\color{red}
The panel (a) shows the reciprocal gating scheme, where ACh molecules do not dissociate until AChRs return to a resting position, and the panel (b) shows the cyclic scheme, where AChRs close after the dissociation of ACh molecules. }
}
\label{fig:gating_scheme} 
\end{figure}

For the competitive kinetic models, we consider the two different schemes for gate opening and closing of AChR as shown in \figref{fig:gating_scheme}. 
In the figure, the complexes formed by binding of ACh, denoted by $\mathrm{A}$, and NDNB, denoted by $\mathrm{D}$, to  AChR,  denoted by $\mathrm{R}$, are represented by 3-letter symbols.  
The first and last letters denote the first and second ligands occupying the sites 1 and 2, respectively, and the middle letter represents the receptor $\mathrm{R}$. 
Unoccupied sites are denoted by $\mathrm{O}$, and $\mathrm{ORO}$ stands for free AChR. 
The parameters $k_{\mathrm{dissA}i}$ and $k_{\mathrm{dissD}i}$ for site $\# i$ $(i =1,\, 2)$ stand for the dissociation rate constants of ACh and NDNB from AChRs, respectively. 
The association constants  $k_{\mathrm{assocA}i}$ and $k_{\mathrm{assocD}i}$ for ACh and NDNB for site $\# i$ are 
given by $k_{\mathrm{assocA}i} := k_{\mathrm{dissA}i}/K_{\mathrm{A}i}$ and $k_{\mathrm{assocD}i} := k_{\mathrm{dissD}i} /K_{\mathrm{D}i}$, respectively, where the parameter $K_{\mathrm{A}i}$ stands for the dissociation equilibrium constant of ACh for site $\# i$. 
In addition, the time courses of gate opening and closing of AChRs are characterized by the rate constants $k_\mathrm{open}$ and $k_\mathrm{close}$, respectively. 
\Figref{fig:reciprocal_scheme} shows the reciprocal gating scheme used in the preceding studies \citep{auerbach98,demazumder01}. 
The symbol ARA stands for AChRs bound with two ACh molecules but in the closed state, and the symbol $\mathrm{ARA}^\ast$ for AChRs in the open state and thus activated due to the conformational change of AChRs. 
On the other hand, in the cyclic scheme shown in \figref{fig:cyclic_scheme}, ACh molecules dissociate before the close of AChR. 
The dissociation and association constants $k_{\mathrm{dissD}i}^\ast$ and $k_{\mathrm{assocA}i}^\ast$ after the activation of AChRs are distinguished from those before the activation. 
In both the two gating  schemes, the symbol $\mathrm{RD}$ represents the desensitized state, and $k_\mathrm{d+}$ and $k_\mathrm{d-}$ stand for the rate constants for desensitization. 
Finally, the decay of the concentration of free ACh molecules in the synaptic cleft, which is mainly due to rapid hydrolysis of ACh by acetylcholinesterase, is characterized by the rate constant $k_\mathrm{decay}$. 
By using the rate constants introduced above, the time course of competition of ACh and NDNB molecules can be described by a set of ordinary differential equations derived based on the framework of chemical kinetics. 
 {\color{red}The model equations and the initial conditions can be found in \cite{nigrovic03,hoshino21} for the reciprocal scheme  and Appendix~A for the cyclic scheme. 
By solving the ordinary differential equation models, the peak concentration of activated AChRs, denoted by $\ce{[R^\ast]}$, can be calculated as the peak concentration of AChRs in open states ($\ce{[ARA^\ast]}$ for the reciprocal scheme and $\ce{[ARA^\ast]}$,  $\ce{[ARO^\ast]}$, and $\ce{[ORA^\ast]}$ for the cyclic scheme).}

%

\begin{table*}[!t]
\caption{ List of parameters in the models of neuromuscular transmission. The symbol ${}^\ast$ stands for the values  reported by \cite{nigrovic03},  ${}^\dagger$ reported by \cite{akk96},  ${}^\ddagger$ reported by \cite{auerbach98}, and  ${}^\S$  reported by \cite{demazumder01}. }
\label{tab:parameters}
\begin{center}
\begin{tabular}{  c | l | c   }
\hline
    symbol  & meaning &  value \\
\hline 
  $\ce{[R]}_\mathrm{total}$ & Concentration of  AChRs in the synaptic cleft & $\SI{7.75e-5}{M}$${}^\ast$\\ 
  $\ce{[A]}_\mathrm{init} $ & {\color{red} Initial concentration of ACh  (\emph{in vivo}) } & {\color{red} $\SI{7.75e-6}{M} $${}^\ast$ } \\ 
                                   & {\color{red} Initial concentration of ACh  (\emph{in vitro}) } &  {\color{red} $  \SI{7.75e-3}{M} $} \\
   $k_\mathrm{decay} $ & {\color{red} Rate constant of the decay of the concentration of free ACh (\emph{in vivo}) }  & {\color{red} $ \SI{1.2e4}{s^{-1}}$${}^\ast$ }  \\
    &  {\color{red} Rate constant of the decay of the concentration of free ACh (\emph{in vitro}) }  & {\color{red} $0$ } \\
  $k_\mathrm{dissA1}$ &  Dissociation rate constant for ACh with site1 of AChR & $ \SI{ 1.8e4}{s^{-1}} $${}^\dag$ \\ 
  $k_\mathrm{dissA2}$ &  Dissociation rate constant for ACh with site2 of  AChR & $ \SI{ 1.8e4}{s^{-1}} $${}^\dag$ \\
   $K_\mathrm{A1}$ &  Dissociation equilibrium constant for ACh with site1 of AChR   & $ \SI{ 1.6e-4 }{ M } $${}^\dag$ \\ 
   $K_\mathrm{A2}$ &  Dissociation equilibrium constant for ACh with site2 of AChR  & $ \SI{ 1.6e-4 }{ M } $${}^\dag$ \\ 
   $k_\mathrm{close}$ &Rate constant of channel closing of AChR &   $\SI{1.2e3}{s^{-1}}$${}^\ddag$  \\ 
  $k_\mathrm{open}$  & Rate constant of  channel opening of AChR  &  $\SI{5.0e4}{s^{-1}}$${}^\ddag$   \\ 
   $k_\mathrm{d+}$    & Rate constant of desensitization  & $\SI{26}{s^{-1}}$${}^\ddag$ \\ 
   $k_\mathrm{d-}$    & Rate constant of recovery from desensitization & $\SI{0.13}{s^{-1}}$${}^\ddag$ \\ 
  $k_\mathrm{dissD1}$ &  Dissociation rate constant for NDNB with site1 of  AChR   & $\SI{12.6}{s^{-1}} $${}^\S$   \\ 
  $k_\mathrm{dissD2}$ &  Dissociation rate constant for NDNB with site2 of AChR  &  $\SI{113}{s^{-1}} $${}^\S$   \\ 
  $K_\mathrm{D1}$ &  Dissociation equilibrium constant for NDNB with site1 of AChR    & $\SI{7.0e-8}{M}${\color{red}${}^\S$}  \\ 
  $K_\mathrm{D2}$ &  Dissociation equilibrium constant for NDNB with site2 of AChR  & $\SI{6.3e-7}{M} ${\color{red}${}^\S$} \\
 $\ce{[R^\ast]_{50}} $ & Concentration of activated AChRs at half-maximal muscle response  &$\SI{9.7e-9}{M}${\color{red}${}^\ast$}  \\
 $\gamma_\mathrm{A} $ &  Slope of the activated AChRs vs  muscle response curve& $4.8${\color{red}${}^\ast$}  \\ 
    \hline 
\end{tabular} 
\end{center}
\end{table*}

After the calculation of the fraction of activated AChRs, the clinically observed muscle response can be simulated in the second step of \figref{fig:model_framework}. 
By using the peak concentration $\ce{[R^\ast]}$ calculated in the first step, the twitch strength is calculated as follows \citep{nigrovic03}: 
\begin{align} \label{eq:twitch_strength}
 \textrm{Twitch Strength} 
 = \dfrac{ \ce{ [R^\ast]}^{\gamma_\mathrm{A}}   }{ \ce{[R^\ast]}^{\gamma_\mathrm{A}} + \ce{[R^\ast]}_\mathrm{50}^{\gamma_\mathrm{A}}  }  
\end{align}
where $\ce{[R^\ast]}_\mathrm{50}$ stands for the parameter representing the concentration of the activated AChRs at half-maximal twitch, and $\gamma_\mathrm{A}$ for the exponent that determines the slope of the sigmoidal curve. 
The formulation of \eref{eq:twitch_strength} is based on the assumption that activation of a defined number of receptors at an end plate triggers the contraction of the associated muscle fiber, and the muscle response is proportional to the number of contracting muscle fibers, while each fiber contracts in an all-or-nothing manner.  

The above models of neuromuscular response can also be used to simulate \emph{in vitro} effects of NDNBs. 
As a typical experimental setting, we postulate a situation where AChRs are expressed into clonal cells, and  outside-out patches are prepared for  voltage-clamp recordings of macroscopic currents (see \cite{wenningmann01,demazumder01}). 
Then, on the condition that the membrane conductance is proportional to the fraction of activated AChRs, the peak current $I_\mathrm{peak}$ after a rapid application of ACh can be described as follows:
\begin{align}
  \dfrac{I_\mathrm{peak}}{I_0} = \dfrac{ \ce{[R^\ast]} }{ \ce{[R^\ast]}_0 }
\end{align}
where $I_0$ stands for the control value of the experimental current in the absence of NDNB.
Furthermore, some simulation settings are changed to consider the \emph{in vitro} environment. 
First, while free ACh molecules are rapidly hydrolyzed by acetylcholinesterase in the synaptic cleft, the concentration of ACh is kept constant in a typical experimental setting. 
Thus, the parameter $k_\mathrm{decay}$ is set to zero for simulating the \emph{in vitro} effects. 
Second, the concentration of ACh used in experiments, represented by the parameter $\ce{[A]_{init}}$, is higher  than that for simulating \emph{in vivo} effects.  
{\color{red}
Specifically, a high concentration is used for \emph{in vitro} experiments such that more than  $\si{95}{\%}$ of AChRs are activated \cite{liu08,liu09}, although a typical number of ACh molecules released \emph{in vivo} is only one tenth of the number of AChRs \citep{salpeter73,hobbiger76}. 
Based on this, we used a concentration as high as $\SI{7.75e-3}{M}$ for \emph{in vitro} simulation such that the above condition holds.  
}

\subsection{Method of Parameter Estimation} \label{sec:parameter_estimation}

\begin{table}[!t]
\caption{In vivo and vitro experimental results used in this study. The values of $\mathrm{EC}_{50}$ and $\gamma_\mathrm{E}$ are reported in \cite{bergeron01}, \cite{fisher97}, and  \cite{plaud95} for cisatracurium, vecuronium, and rocuronium, respectively, and the values of $\mathrm{IC}_\mathrm{50}$ and $\gamma_\mathrm{I}$ reported in  \cite{liu08}. 
} \label{tab:pharmacologic_data}     
\begin{tabular}{ l  l  l c   c  c  }
\hline\noalign{\smallskip}
 &  & Cisatracurium  & Vecuronium & Rocuronium  \\
\noalign{\smallskip} \hline \noalign{\smallskip} 
 \multicolumn{3}{ l }{In Vivo results} \\ \noalign{\smallskip}
& $\mathrm{EC}_{50} \, (\si{\micro M})$ &   $0.12 \pm 0.027$ &  $0.26 \pm 0.10 $ & $ 1.35 \pm 0.26$  \\
& $\gamma_\mathrm{E}$  &  $6.9 \pm 1.3 $  &  $7.6 \pm 3.8$ & $ 4.79 \pm 1.70 $  \\
\noalign{\smallskip}
 \multicolumn{3}{ l }{In Vitro results } \\ \noalign{\smallskip}
& $\mathrm{IC}_{50} \, ( \si{\nano M})  $  & $10 \pm 1$ &          $15 \pm 2 $ & $ 17 \pm 2 $  \\
& $\gamma_\mathrm{I} $  & $1.02 \pm 0.09 $ &  $1.03 \pm 0.12 $ & $ 0.67 \pm 0.05 $ \\
\noalign{\smallskip}\hline
\end{tabular}
\end{table}

In this paper, a set of  estimates of parameters for each model is determined based on \emph{in vivo} and \emph{in vitro} experimental results reported in literature. 
The clinical effects of NDNBs have been quantified based on pharmacokinetic and pharmacodynamic (PKPD) analyses, and the relationship between the estimates of concentration of NDNBs in the effect compartment and the twitch strength is fitted to the so-called Hill equation \citep{sheiner79}. 
As a result, the values of $\mathrm{EC}_{50}$, the concentration needed to produce a $\SI{50}{\%}$ decrease of muscle response,  and $\gamma_\mathrm{E}$, the Hill coefficient, are reported \cite{bergeron01, fisher97, plaud95} as listed in \tabref{tab:pharmacologic_data}. 
While other studies have also been reported, we selected these studies considering that data were obtained from patients under propofol anesthesia rather than isoflurane anesthesia and that neuromuscular monitoring was performed using  a mechanomyography (force transducer) rather than acceleromyography or electromyography to utilize experimental results obtained under similar conditions. 
Similarly, \emph{in vitro} effects of NDNBs have also been fitted by the Hill equation, and  the values of $\mathrm{IC}_\mathrm{50}$, the concentrations needed to produce a $50\%$ inhibition of the current, and $\gamma_\mathrm{I}$, the associated Hill coefficient, are reported \citep{liu09}. 
Although some rate constants have been measured and reported \cite{wenningmann01,demazumder01,demazumder08}, these studies have used mouse AChRs. 
This paper determines these rate constants using the values of $\mathrm{IC}_\mathrm{50}$ and $\gamma_\mathrm{I}$ reported in  \cite{liu08} obtained by using human adult AChRs rather than mouse adult or embryonic AChRs.

{\color{red}
Besides the kinetic constants for each of the three NDNBs (such as $k_\mathrm{dissD1}$ and $K_\mathrm{D1}$), we estimate  the kinetic constants for ACh (such as $k_\mathrm{dissA1}$ and $k_\mathrm{dissA2}$) as well. 
While these constants have been reported in \cite{akk96}, these values were estimated for mouse adult AChR and not necessarily applicable for human adult AChR.
Also, the parameters $\ce{[R^\ast]}_\mathrm{50}$ and $\gamma_\mathrm{A}$ have been calculated considering a hypothetical drug and not from experimental data of clinically used NDNBs.  
Thus, we re-estimate these parameters.}
The problem of parameter estimation can be formulated as an optimization problem with the following objective function $F$:
\begin{align} \label{eq:cost_function}
 F   = & \dfrac{ 1 }{ 4N_\mathrm{D} } \sum_{ k =1 }^{N_\mathrm{D}}  
         \left \{   
             \dfrac{ (S_{\mathrm{EC}_{50},k} - E_{\mathrm{EC}_{50}, k} )^2 }{ \mathrm{CI}_{\mathrm{EC}_{50}, k}^2 }
            +  \dfrac{ (S_{\gamma_\mathrm{E},k} - E_{\gamma_\mathrm{E}, k} )^2 }{ \mathrm{CI}_{\gamma_\mathrm{E},k}^2 }
            \right. \notag \\
           & \left. +  \dfrac{ (S_{\mathrm{IC}_{50},k} - E_{\mathrm{IC}_{50},k} )^2 }{ \mathrm{CI}_{\mathrm{IC}_{50},k}^2 }
            +  \dfrac{ (S_{\gamma_\mathrm{I},k} - E_{\gamma_\mathrm{I}, k} )^2 }{ \mathrm{CI}_{\gamma_\mathrm{I},k}^2 }
       \right \}
       \notag \\ 
       & + 
       \dfrac{W}{4} \sum_{i=1}^2 \left\{ ( \log_{10} k_{\mathrm{dissA}i}^\mathrm{est} / k_{\mathrm{dissA}i}^\mathrm{nom} )^2 + ( \log_{10} k_{\mathrm{assocA}i}^\mathrm{est} / k_{\mathrm{assocA}i}^\mathrm{nom} )^2 \right\} 
\end{align}
where $k$ represents the index for each NDNB (1: cisatracurium, 2: vecuronium, and 3: rocuronium in this paper), and $N_\mathrm{D}$ stands for the number of NDNBs considered ($N_\mathrm{D}=3$ in this paper). 
The symbol $S$ stands for the simulated value of the pharmacologic parameter indicated by the subscript $\mathrm{EC}_{50}$, $\gamma_\mathrm{E}$, $\mathrm{IC}_{50}$, or $\gamma_\mathrm{I}$, and the symbol $E$ for the experimental results. 
The symbol $\mathrm{CI}$ stands for the  $95 \%$ confidence interval of the experimental result to normalize the errors between experimental and simulation results. 
The second term of $F$ represents the penalty term for the difference between estimated and nominal values of the dissociation and association constants $k_{\mathrm{dissA}i}$ and $k_{\mathrm{assocA}i}$  
shown in \tabref{tab:parameters}.
Furthermore, considering that the two binding sites of mouse adult AChR have similar affinities for ACh \citep{akk96},
we assume that it is also the case for human adult AChR and estimate the parameters $k_\mathrm{dissA}$ and $K_\mathrm{A}$ with $k_\mathrm{dissA}=k_{\mathrm{dissA}1}=k_{\mathrm{dissA}2}$ and  $K_\mathrm{A}= K_\mathrm{A1}= K_\mathrm{A2}$. 
{\color{red}With the aim of reducing a number of estimated parameters,  we also assume that the dissociation rate constants $k_{\mathrm{dissD}1}$ and  $k_{\mathrm{dissD}2}$ are equal ($k_\mathrm{dissD}=k_{\mathrm{dissD}1}=k_{\mathrm{dissD}2}$).}

The proposed framework of parameter estimation was implemented in python codes. 
To numerically solve ordinary differential equations of the models, the Fortran-based solver LSODA provided by the python package SciPy (Version 1.5.2) was used, and the time courses of competitive kinetics were simulated for $\SI{5}{ms}$.
For the nonlinear regression analysis to derive the pharmacologic parameters (such as $\mathrm{EC}_{50}$ and $\mathrm{IC}_{50}$) from a calculated concentration-effect curve, a trust region reflective algorithm implemented in the \verb|least_square| function provided by the package SciPy was used. 
Finally, the cost function $F$ was minimized by Nelder-Mead Algorithm implemented in the \verb|minimize| function in the same package.

\section{Results} \label{sec:results}

\begin{table}[!t]
\caption{Results of parameter estimation for the three modeling structures}
\label{tab:results}    
\begin{tabular}{ l   c  c  c  c  }
\hline\noalign{\smallskip}
  & (a) & (b) & (c) \\ 
  Parameters &  Two-site binding model & Competitive kinetic model & Competitive kinetic model   \\ 
  & &  (Reciprocal scheme)  & (Cyclic scheme) \\
\noalign{\smallskip}\hline \noalign{\smallskip} 
$F$                   & 4.88 & {\color{red}1.19} & {\color{red}0.62} \\
1st term of $F$  & 4.88 & 0.14 & 0.16 \\
2nd term of $F$ & ---  & {\color{red}1.05}  & {\color{red}0.46} \\
\noalign{\smallskip}\hline \noalign{\smallskip} 
  {\color{red} $\ce{ [R^\ast]}_\mathrm{50}$}  &  $ ( 1.33\times 10^{-6} )\ce{[R^\ast]_0}$ & $\SI{ 2.08e-07 }{M}$   &  $\SI{ 1.82 e-8 }{M}$ \\
  $\gamma_A$              & $4.17$& $9.14$ & $9.04$ \smallskip  \\ 
  $k_\mathrm{dissA}$    & ---    &  $\SI{ 4.43e+02 }{s^{-1}}$      &  $\SI{ 2.62 e3 }{s^{-1}}$ \\
  $K_\mathrm{A}$         & ---    &  $\SI{ 1.58e-8}{M}$               &  $\SI{ 4.44e-7}{M} $  \\ 
  $k_\mathrm{close}$    &---     &   $\SI{ 2.22e7 }{s^{-1}}$         &  $\SI{ 4.24e4 }{s^{-1}}$ \\
  $k_\mathrm{open}$     & ---    &  $\SI{ 1.48e10 }{s^{-1}}$          &  $\SI{ 1.06e4 }{s^{-1}}$ \\
  $k_\mathrm{dissA}^\ast$  &--- & ---                                   & $\SI{ 1.70e4 }{s^{-1}}$ &  \\
  $K_\mathrm{A}^\ast$  & --- & ---                                       & $\SI{ 1.84e-8 }{M}$ & \\
\noalign{\smallskip}\hline\noalign{\smallskip} 
  \underline{Cisatracrium} \smallskip \\
  $ K_\mathrm{D1}$    & $\SI{2.19 e-8}{M}$  &  $\SI{9.57e-9}{M}$ & $\SI{1.02e-08}{M}$  \\
  $ K_\mathrm{D2}$    & $\SI{2.12 e-8}{M}$  &  $\SI{6.75e-6}{M}$ & $\SI{3.60e-05}{M}$ \\
  $ k_\mathrm{dissD}$ &  ---                      &  $\SI{2.6}{s^{-1}}$    & $\SI{4.0}{s^{-1}}$ \smallskip \\
  \underline{Vecuronium} \smallskip \\
  $ K_\mathrm{D1}$    & $\SI{2.09 e-8}{M}$  & $\SI{1.58e-08}{M}$ & $\SI{1.63e-08}{M}$\\ 
  $ K_\mathrm{D2}$    & $\SI{9.29 e-8}{M}$  & $\SI{2.76e-06}{M}$ & $\SI{1.58e-06}{M}$ \\
  $ k_\mathrm{dissD}$ &  ---                      & $\SI{1.9}{s^{-1}}$    & $\SI{5.4}{s^{-1}}$  \smallskip \\
  \underline{Rocuronium} \smallskip \\
  $ K_\mathrm{D1}$    & \SI{1.88 e-8}{M}     & $\SI{1.23e-08}{M}$  &$\SI{1.22e-08}{M}$\\
  $ K_\mathrm{D2}$    & \SI{1.28 e-4}{M}     & $\SI{1.37e-07}{M}$ & $\SI{1.76e-07}{M}$\\
  $ k_\mathrm{dissD}$ &  ---                     & $\SI{64.0}{s^{-1}}$   & $\SI{61.5}{s^{-1}}$ \\
\noalign{\smallskip}\hline
\end{tabular}
\end{table}

\begin{figure}[!t]
\centering
 \begin{minipage}[t]{0.44\linewidth} \centering
  \includegraphics[width=0.9\linewidth]{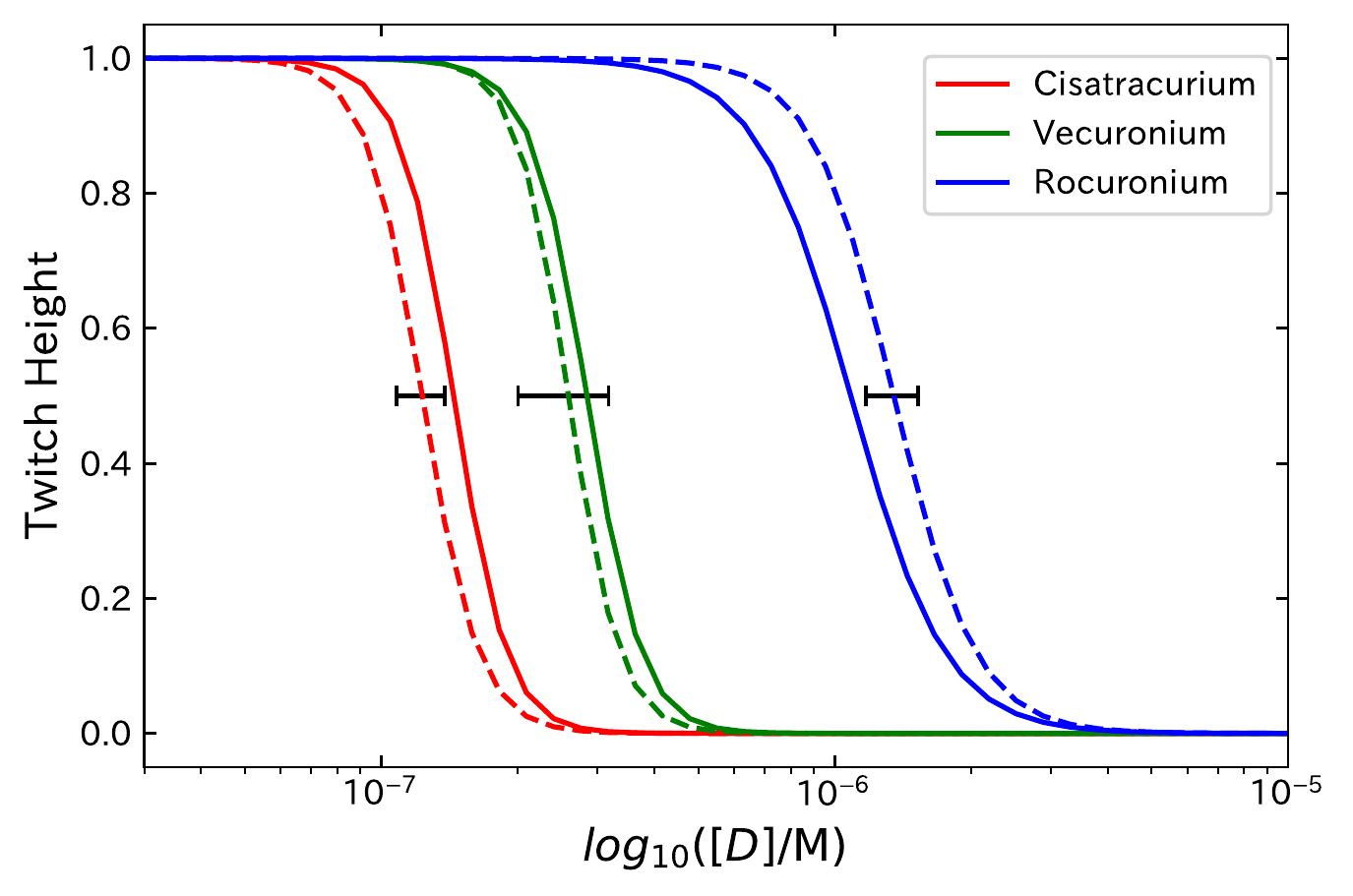}
  \subcaption{Two-site binding model (\emph{in vivo})} \label{fig:two-site-invivo}
 \end{minipage}
 \begin{minipage}[t]{0.44\linewidth} \centering
  \includegraphics[width=0.9\linewidth]{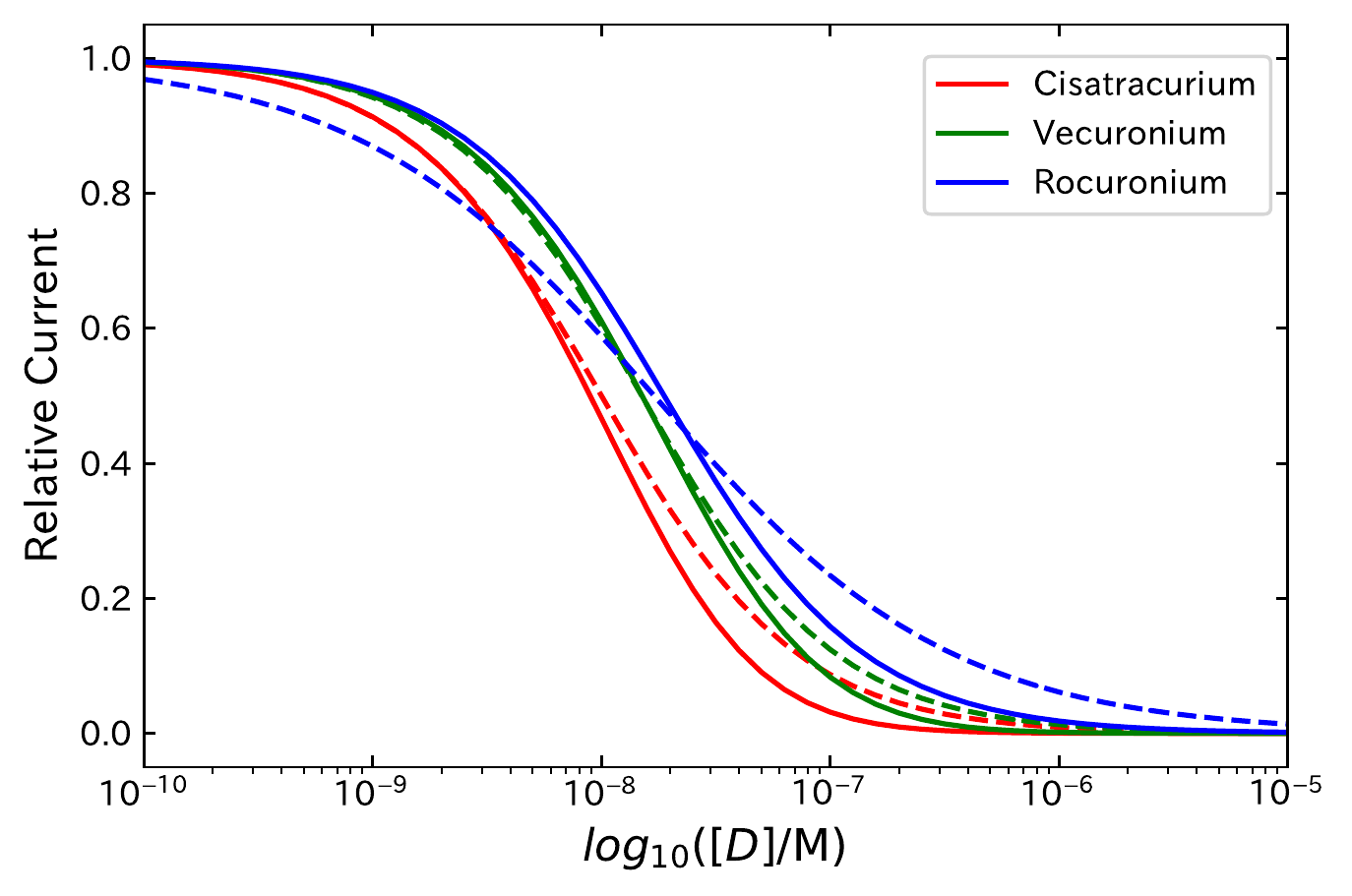}
  \subcaption{Two-site binding model (\emph{in vitro})} \label{fig:two-site-invitro}
 \end{minipage} \\[2mm]
 \begin{minipage}[t]{0.44\linewidth}  \centering
  \includegraphics[width=0.9\linewidth]{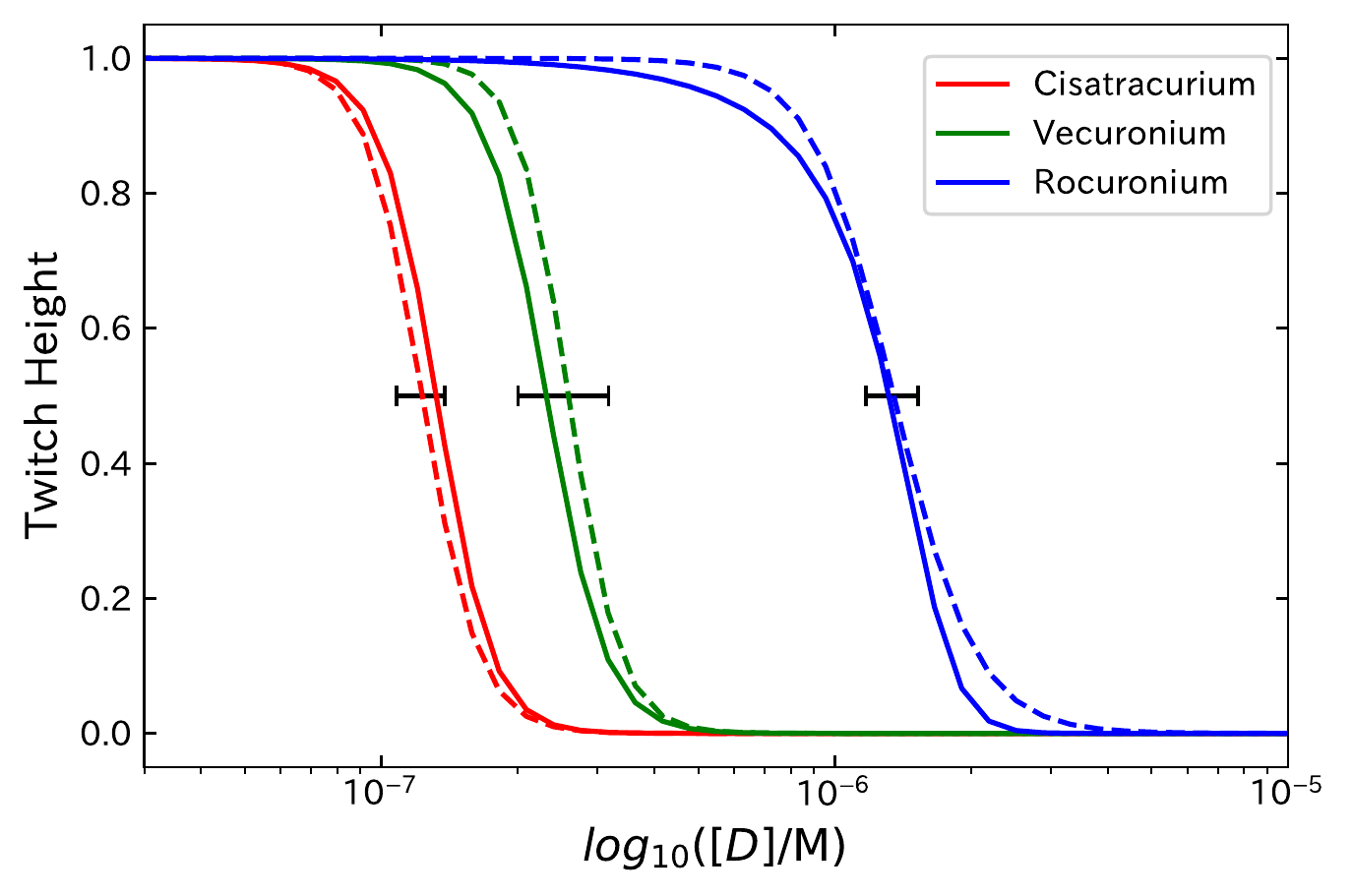}
  \subcaption{Reciprocal scheme (\emph{in vivo})} \label{fig:reciprocal-invivo}
  \end{minipage}
 \begin{minipage}[t]{0.44\linewidth}  \centering 
  \includegraphics[width=0.9\linewidth]{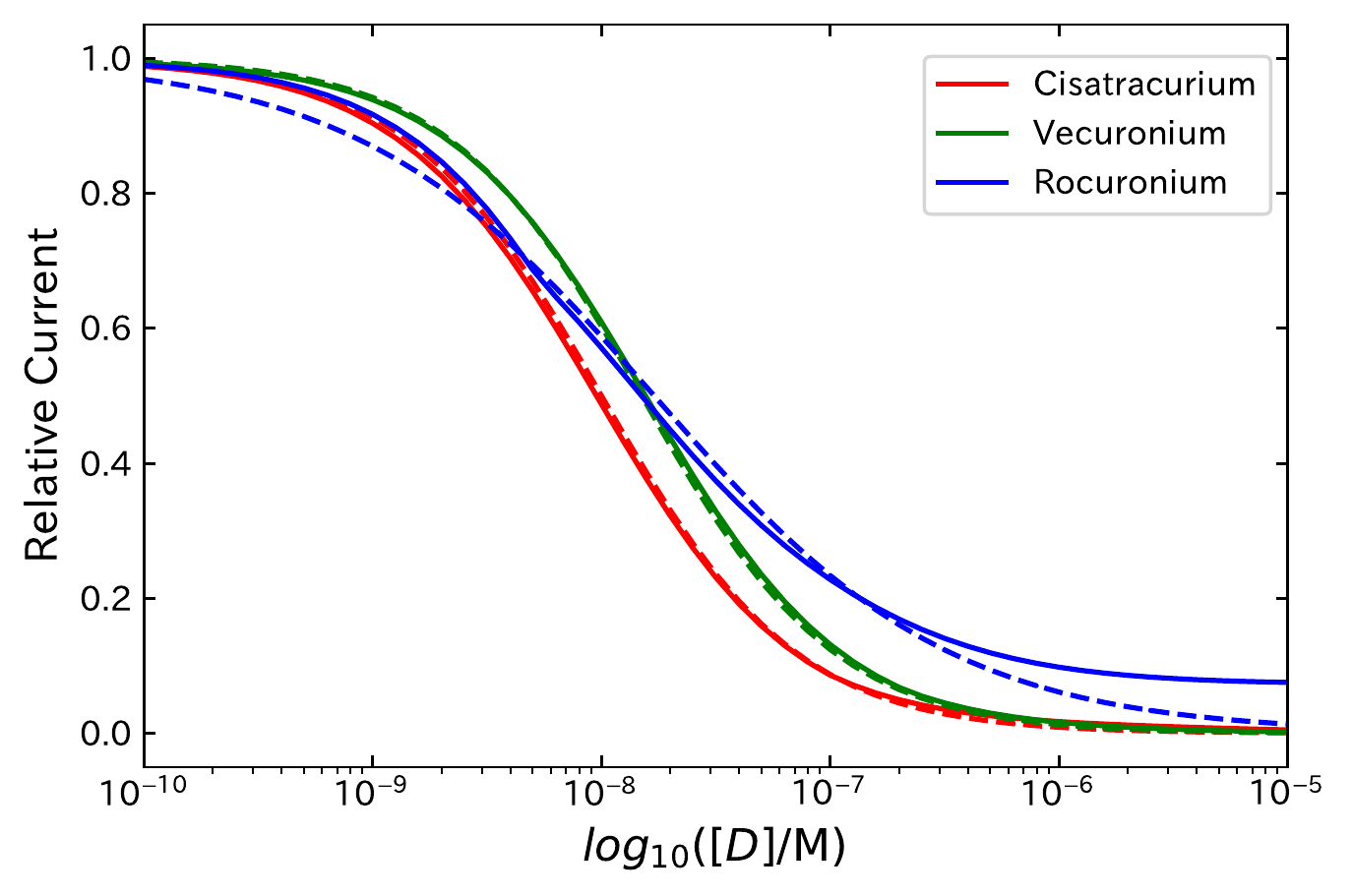}
  \subcaption{Reciprocal scheme (\emph{in vitro})}  \label{fig:reciprocal-invitro}
  \end{minipage}    
 \begin{minipage}[t]{0.44\linewidth} \centering 
  \includegraphics[width=0.9\linewidth]{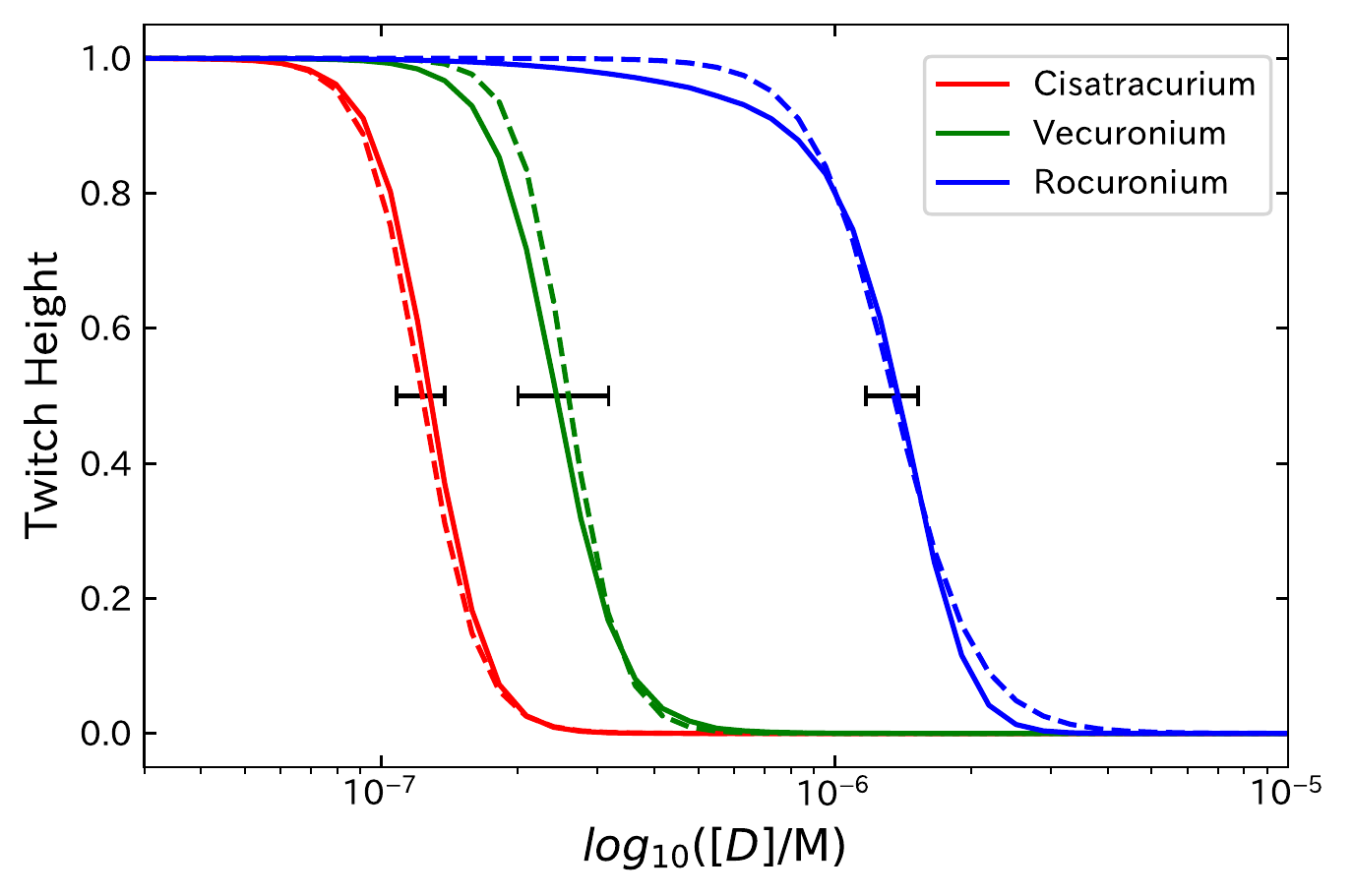}
  \subcaption{Cyclic scheme (\emph{in vivo})}  \label{fig:cyclic-invivo}
  \end{minipage}
 \begin{minipage}[t]{0.44\linewidth}  \centering 
  \includegraphics[width=0.9\linewidth]{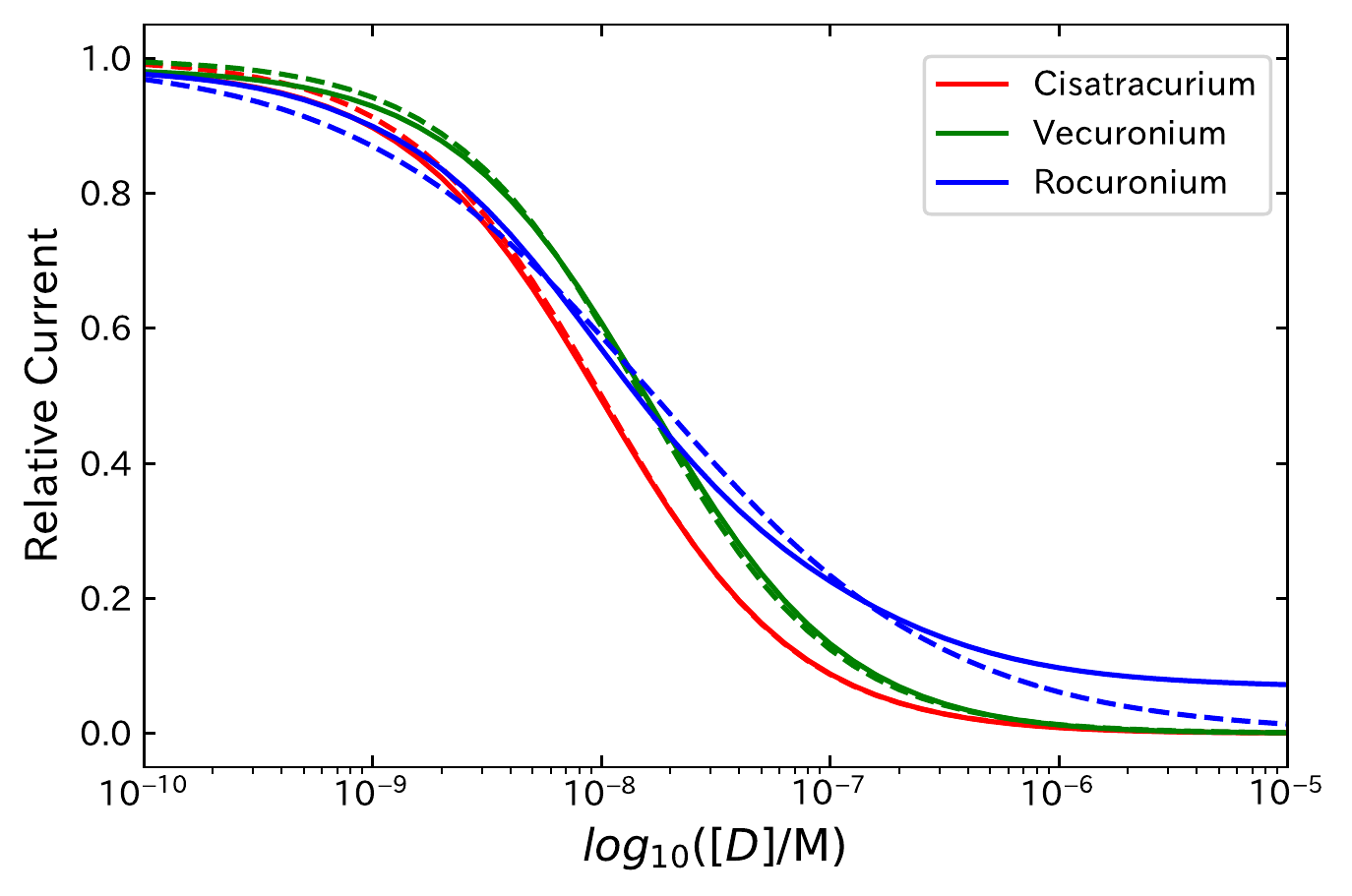}
  \subcaption{Cyclic scheme (\emph{in vitro})} \label{fig:cyclic-invitro}
  \end{minipage}    
\caption{Concentration-effect relationship for the two-site binding model (a, b), competitive kinetic model with reciprocal gating scheme (c,d), and competitive kinetic model with cyclic gating scheme (e,f). The solid lines show the simulation results, and the broken lines the sigmoidal curves plotted based on the experimental results given in \tabref{tab:pharmacologic_data}.  }
\label{fig:concentration-effect-relationship}
\end{figure}

The results of parameter estimation are summarized in \tabref{tab:results}. 
For the two-site binding model, the eight parameters shown in the table, including the dissociation equilibrium constants $K_\mathrm{D1}$ and $K_\mathrm{D2}$ for each NDNB, were estimated. 
The value of the objective function $F=4.88$ is larger than $1$, implying  that the averaged error between simulation and experimental results is larger than the $95 \%$ confidence interval of the experimental results. 
\Figrefs{fig:two-site-invivo}{fig:two-site-invitro} show the comparison between simulation and experimental results for \emph{in vivo} and \emph{in vitro} effects, respectively.  
The solid lines in the figures show simulation results, and the broken lines the sigmoidal curves based on the experimental results given in \tabref{tab:pharmacologic_data}. 
The error bars in \figref{fig:two-site-invivo} show the $95\%$ confidence intervals of  $\mathrm{EC}_{50}$ for each NDNB, and it can be seen that the simulated values of $\textrm{EC}_{50}$ for cisatracurium and rocuronium are out of the confidence interval. 
This clarifies that the two-site binding model is insufficient for the consistent modeling of \emph{in vivo} and \emph{in vitro} effects.

\begin{figure}[!t]
 \centering
 \begin{minipage}[t]{0.45\linewidth}
  \includegraphics[width=0.95\linewidth]{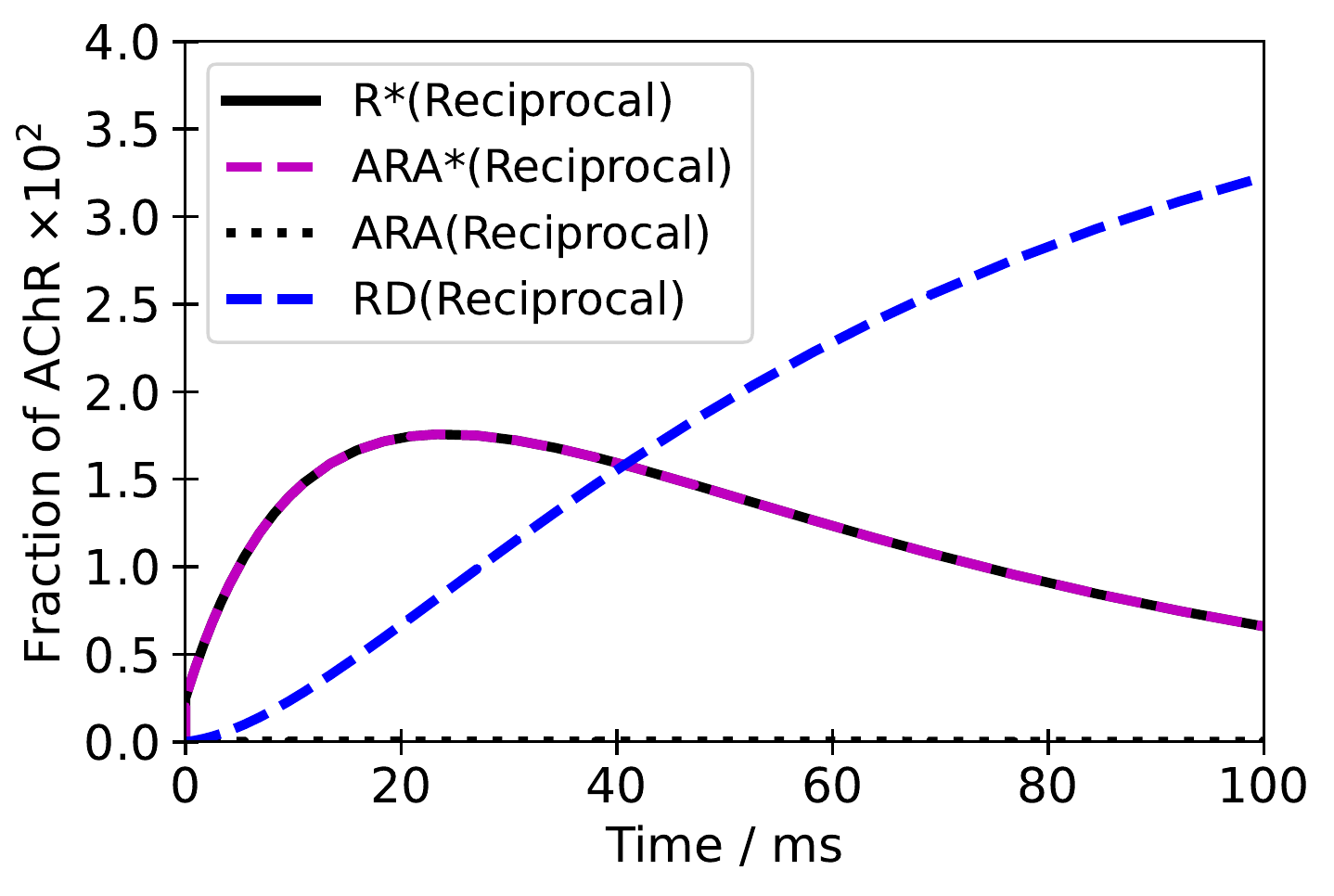}
  \subcaption{Reciprocal scheme} \label{fig:reciprocal_time_course}
  \end{minipage}
 \begin{minipage}[t]{0.45\linewidth}
  \includegraphics[width=0.95\linewidth]{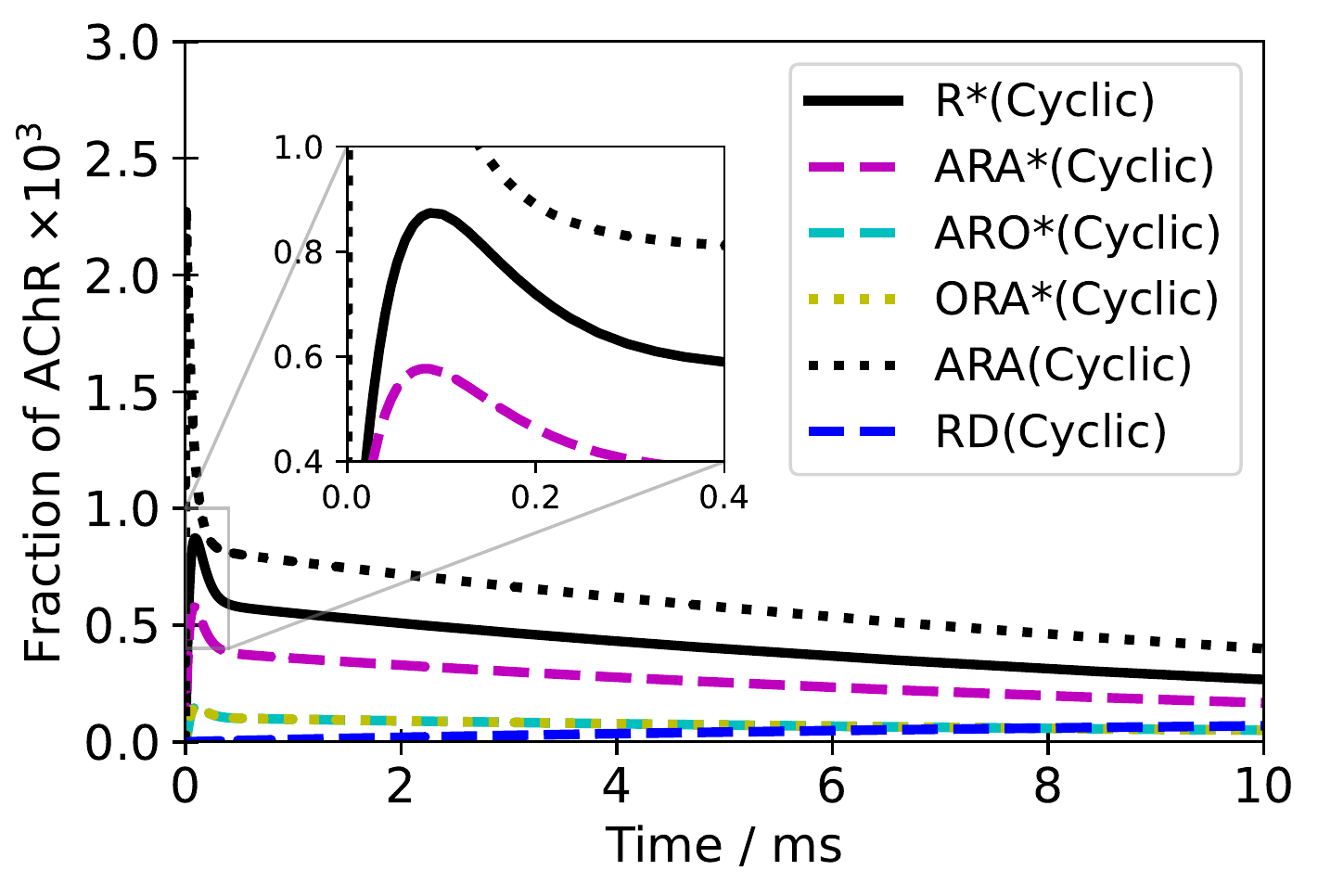}
   \subcaption{Cyclic scheme} \label{fig:cyclic_time_course}
 \end{minipage}
 \caption{\color{red}The time course of activation of AChRs simulated by (a) the reciprocal scheme and (b) the cyclic scheme, under the setting of $[D]=0$. The reciprocal scheme shows unrealistically slow time course such that desensitization takes place, which do not occur \emph{in vivo} with a single activation event. }
 \label{fig:time_course}       
\end{figure}

%
{\color{red}
The competitive kinetic models have a high descriptive ability such that the first term of the objective function $F$, which quantifies the error between simulated and experimental results, can be small (0.14 and 0.16 for reciprocal and cyclic schemes, respectively). 
To obtain this result, we chose the weight of the penalty term in \eqref{eq:cost_function} as $W = 0.25$. 
This is chosen such that the first term of $F$ remains small and the simulated concentration-effect curves are close to the experimental results as shown in \figrefs{fig:reciprocal-invivo}{fig:reciprocal-invitro} for the reciprocal scheme and  \figrefs{fig:cyclic-invivo}{fig:cyclic-invitro} for the cyclic scheme. 
The resultant $\mathrm{EC}_{50}$ values lie in their confidence intervals. 
On the other hand, we can see that there is a 
significant difference between the estimated results for the reciprocal and cyclic scheme in the second term of the objective function $F$ (1.05 for reciprocal scheme and 0.46 for cyclic scheme). 
This means that the dissociation rate constant $k_\mathrm{dissA}$ and the dissociation equilibrium constant $K_\mathrm{A} = k_\mathrm{dissA}/k_\mathrm{assocA}$ of the cyclic scheme are closer to the nominal values than those of the reciprocal scheme. 
These parameters affect the time course of activation of AChRs as shown in \figref{fig:time_course}. 
With the reciprocal scheme, it can be seen that the time course of $\ce{[R^\ast]}$ is much slower than the cyclic scheme, and  it does not reach its peak concentration within initial activation phase and that it has peak at around several tens of $\si{ms}$. 
This is due to the slow dissociation ($k_\mathrm{dissA}=\SI{4.43e2}{ s^{-1} }$) and the high affinity ($K_\mathrm{A} = \SI{1.58e-8}{M}$) of ACh to AChRs. 
This time course for the reciprocal model  is too slow, given that an end-plate current has a typical time constant of about $\si{ms}$ and desensitization will not occur \emph{in vivo} with a single activation event \cite{auerbach98}.
In contrast, with the cyclic scheme, $\ce{[R^\ast]}$ reaches its peak in less than $\SI{1}{ms}$ and desensitization is not observed. 
Thus, we conclude that the cyclic scheme is more appropriate than the reciprocal scheme.}

To better understand the difference among the descriptive abilities of the three model structures, here we analyze how the pharmacologic parameters $\mathrm{EC}_{50}$, $\gamma_\mathrm{E}$, $\mathrm{IC}_{50}$, and $\gamma_\mathrm{I}$ change depending on the parameters $K_\mathrm{D1}$, $K_\mathrm{D2}$, and $k_\mathrm{dissD}$ characterizing the properties of an NDNB. 
For this analysis, it is convenient to utilize the fact that the models can be represented in a dimensionless normalized form (see \cite{hoshino21}), where the concentrations $\mathrm{EC}_{50}$ and $\mathrm{IC}_{50}$ are normalized by the dissociation equilibrium constant $K_{\mathrm{D}1}$, and the properties of an NDNB can be identified by the ratio of the two  dissociation equilibrium constants $K_{\mathrm{D}1}$ and $K_{\mathrm{D}2}$, i.e. the site-selectivity $\mu := K_{\mathrm{D}2}/K_{\mathrm{D}1}$, as well as the dissociation rate constants $k_\mathrm{dissD}$.

%
\begin{figure}[!t]
 \centering
 \begin{minipage}[t]{0.40\linewidth}
  \includegraphics[width=0.8\linewidth]{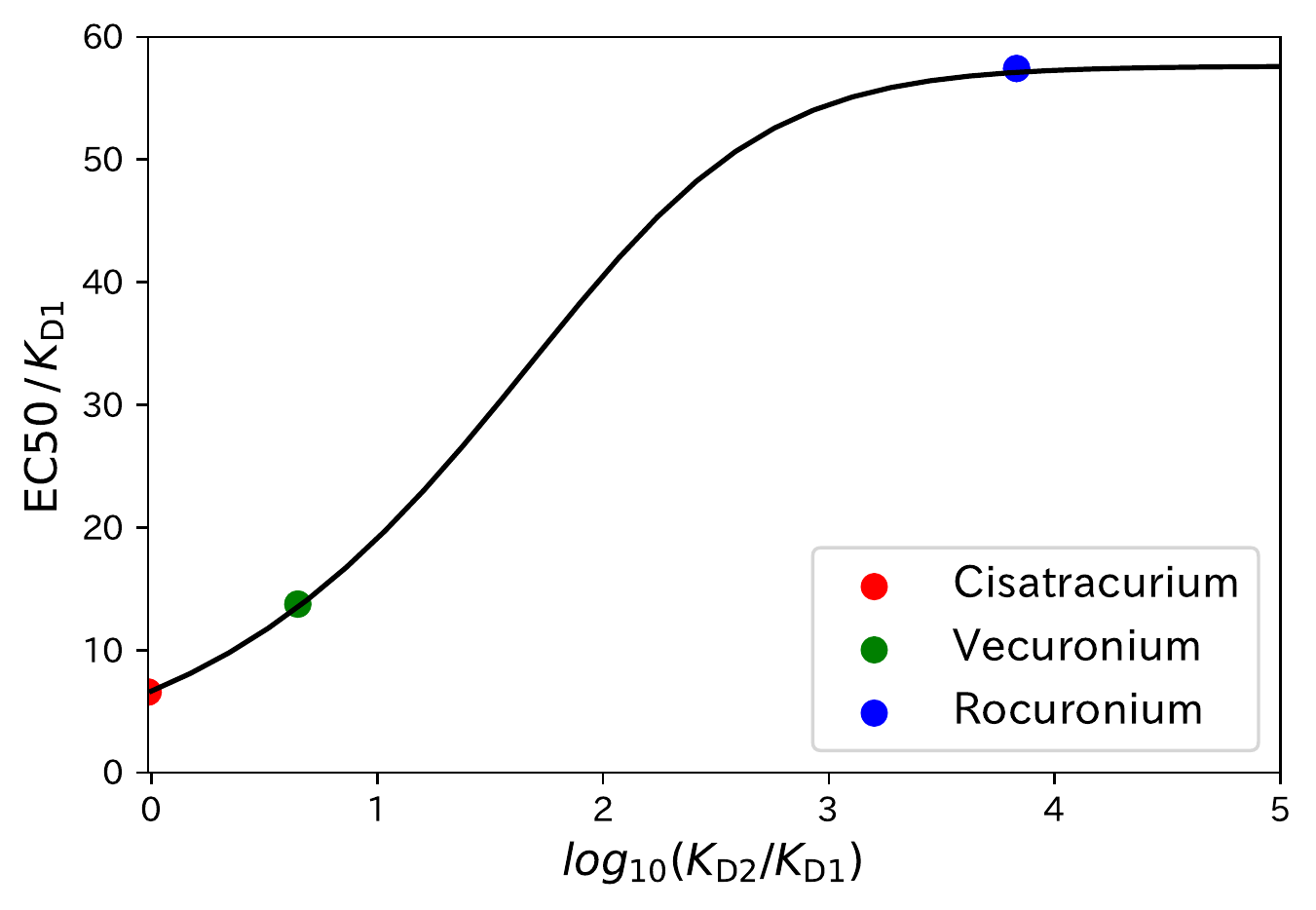}
  \subcaption{$\textrm{EC}_{50}$} \label{fig:two-site_EC50}
  \end{minipage}
 \begin{minipage}[t]{0.40\linewidth}
  \includegraphics[width=0.8\linewidth]{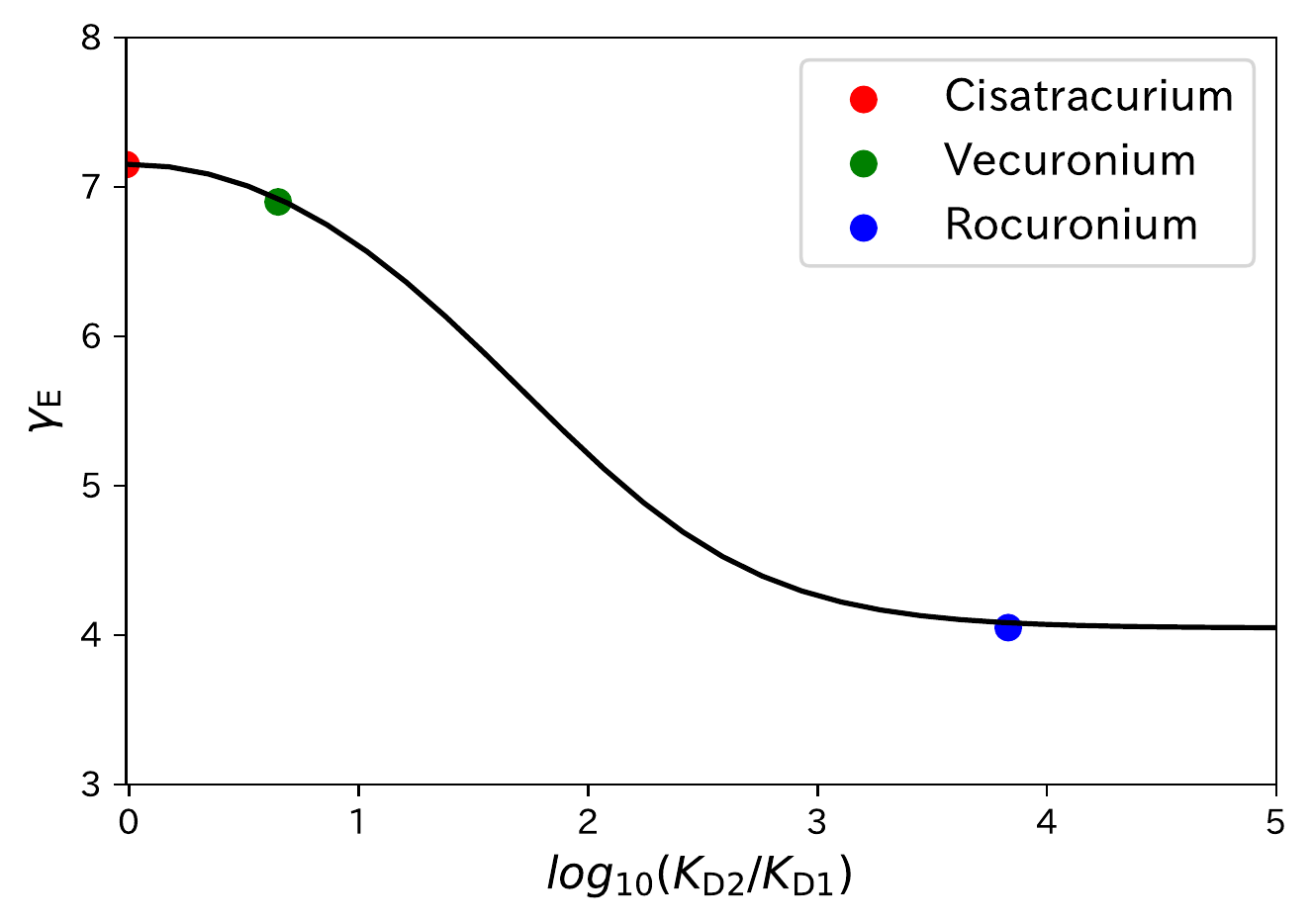}
   \subcaption{$\gamma_\mathrm{E}$} \label{fig:two-site_gammaE}
 \end{minipage}
 \begin{minipage}[t]{0.40\linewidth}
  \includegraphics[width=0.8\linewidth]{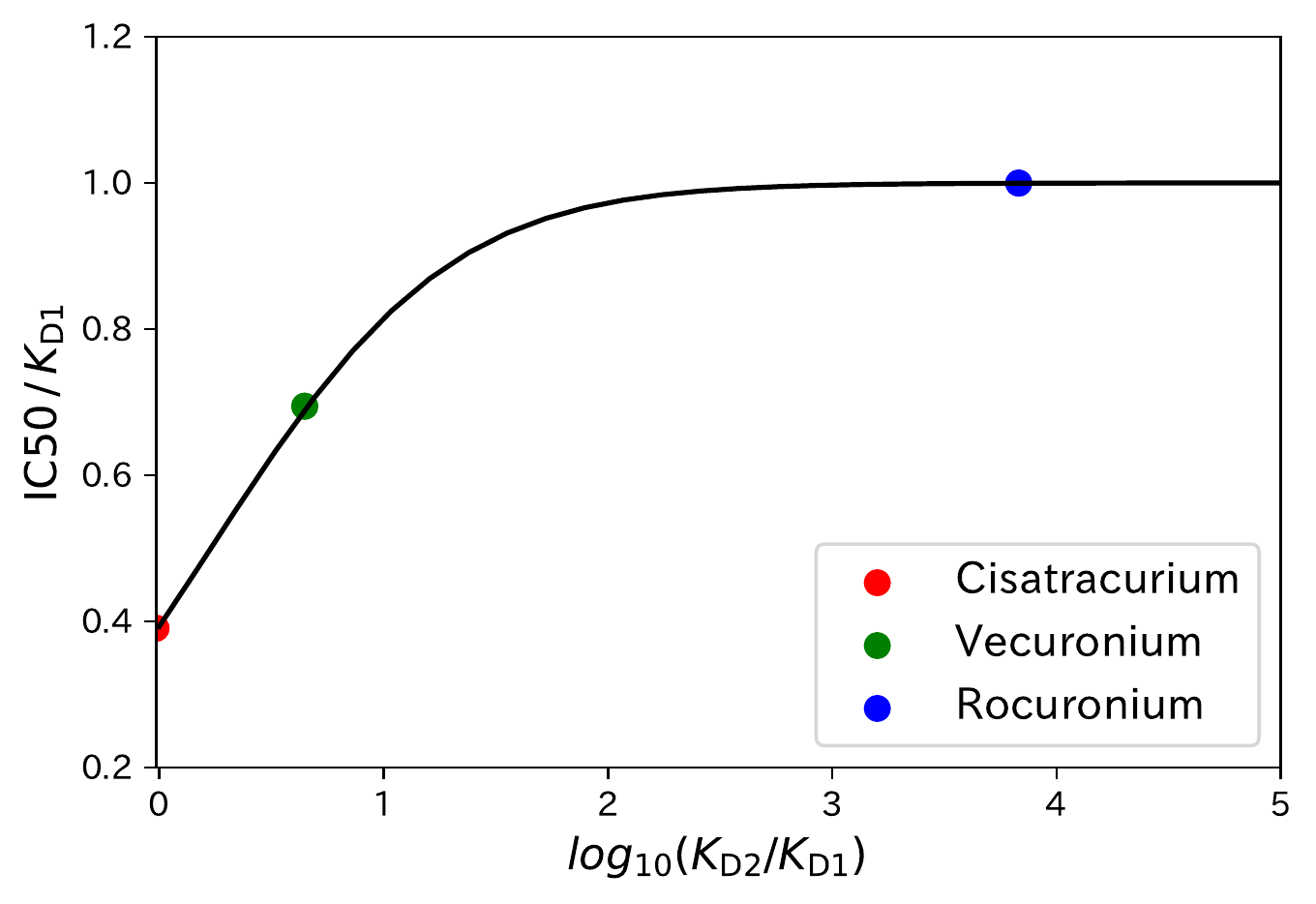}
   \subcaption{$\textrm{IC}_{50}$}  \label{fig:two-site_IC50}
 \end{minipage}
 \begin{minipage}[t]{0.40\linewidth}
  \includegraphics[width=0.8\linewidth]{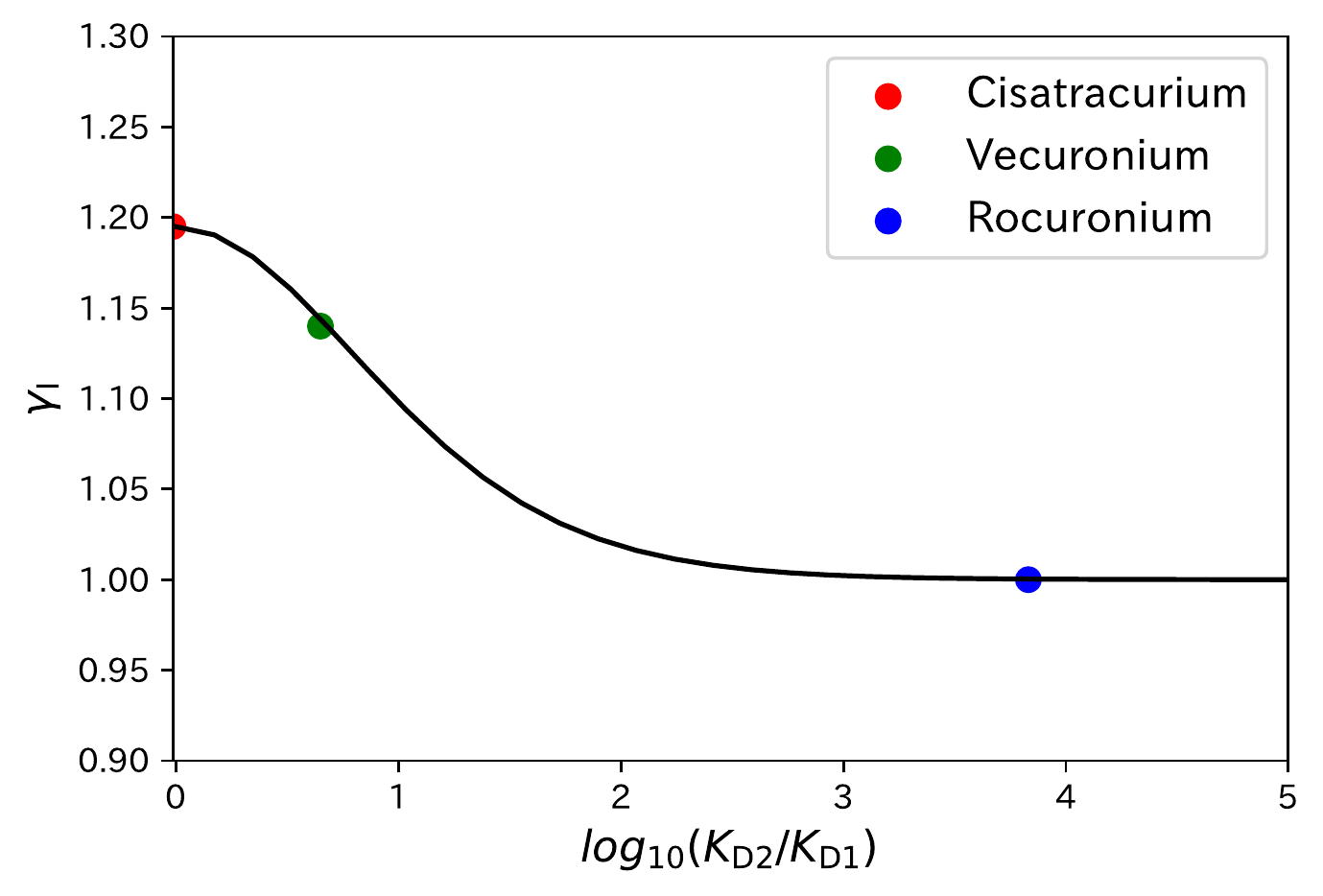}
   \subcaption{$\gamma_\mathrm{I}$} \label{fig:two-site_gammaI}
 \end{minipage}
 \caption{Pharmacologic parameters simulated by the two-site binding model with different values of  $\mu := K_\mathrm{D2}/K_\mathrm{D1}$. The \emph{red}, \emph{green}, and \emph{blue} points show the calculated values for cisatracurium, vecuronium, and rocuronium, respectively.}
 \label{fig:trend_two-site_model}       
\end{figure}
\Figref{fig:trend_two-site_model} shows the results of pharmacologic parameters simulated by the two-site binding model with different values of  $\mu$ (the parameter $k_\mathrm{dissD}$ is not used in this model). 
The values of $\mathrm{EC}_{50}$ and $\mathrm{IC}_{50}$ monotonically increase with an increase of the parameter $\mu$, and the values of  $\gamma_\mathrm{E}$ and $\gamma_\mathrm{I}$ decrease with an increase of  the parameter $\mu$. 
It can be seen that the value of $\gamma_\mathrm{I}$ is in the range of $[1.0, 1.2]$, and thus the experimental result of $\gamma_\mathrm{I}$ for rocuronium ($\gamma_\mathrm{I} = 0.67 \pm 0.05$) cannot be well described by the two-site binding model. 
The red, green, and blue points in the figure show the results for the cases of cisatracurium, vecuronium, and rocuronium, respectively. 
It can be seen that the site selectivity $\mu$ is almost 1 for cisatracurium, and takes a large value for rocuronium ($\mu = 6.8\times10^3$). 

\begin{figure}[t]
 \centering
 \begin{minipage}[t]{0.40\linewidth}
  \includegraphics[width=0.75\linewidth]{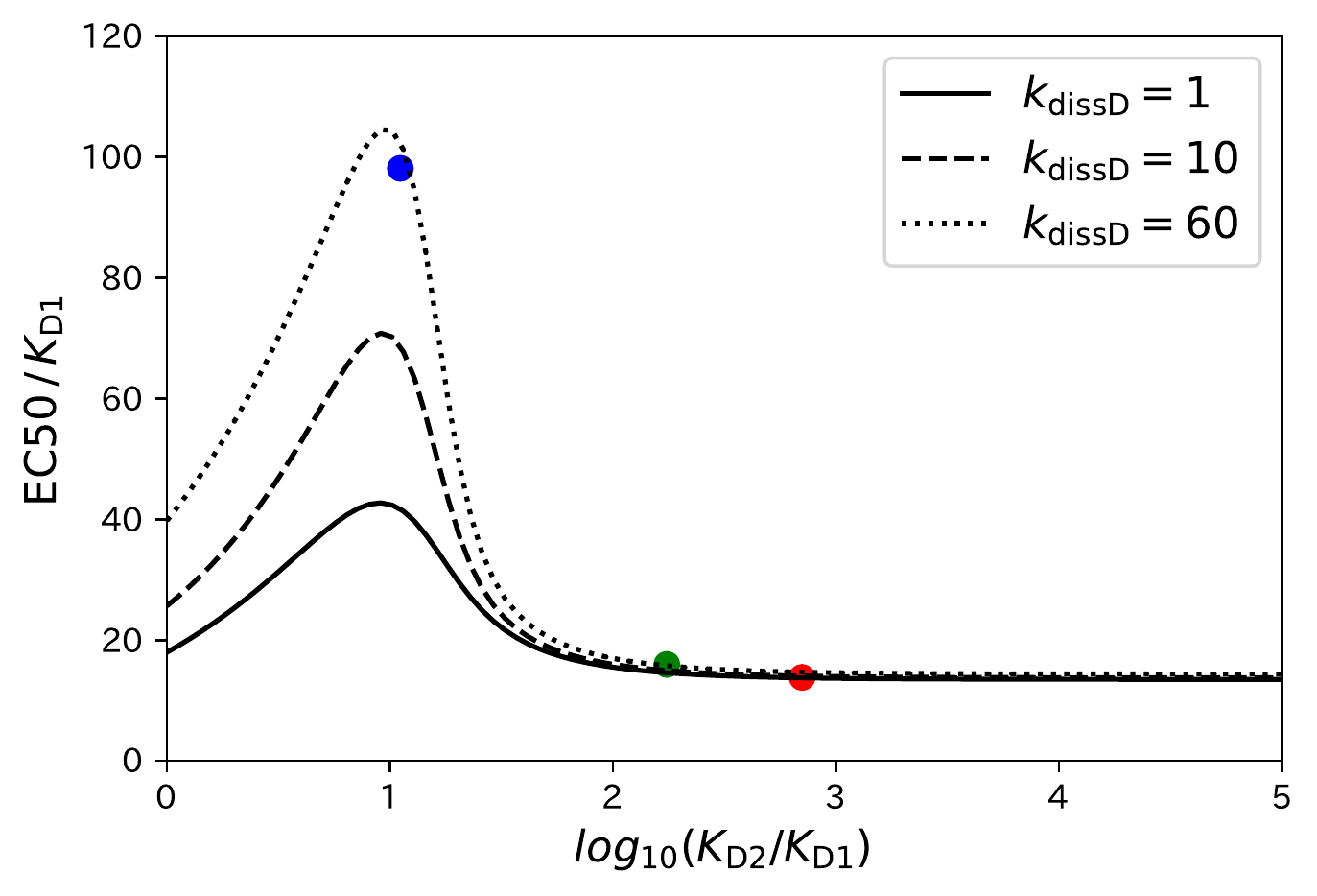}
  \subcaption{$\textrm{EC}_{50}$} \label{fig:reciprocal_EC50}
  \end{minipage}
 \begin{minipage}[t]{0.40\linewidth}
  \includegraphics[width=0.75\linewidth]{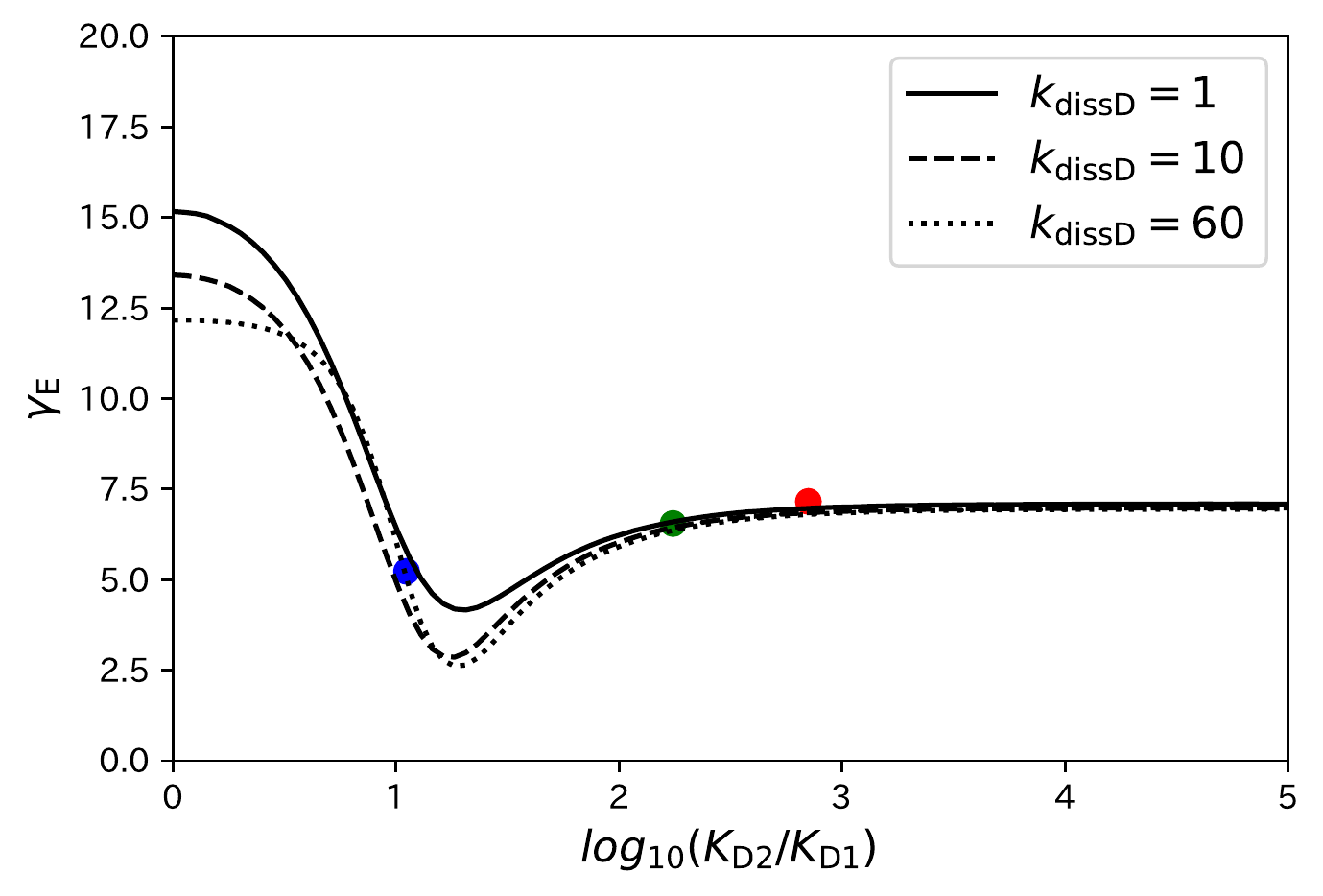}
   \subcaption{$\gamma_{E}$} \label{fig:reciprocal_gammaE}
 \end{minipage}
 \begin{minipage}[t]{0.40\linewidth}
  \includegraphics[width=0.75\linewidth]{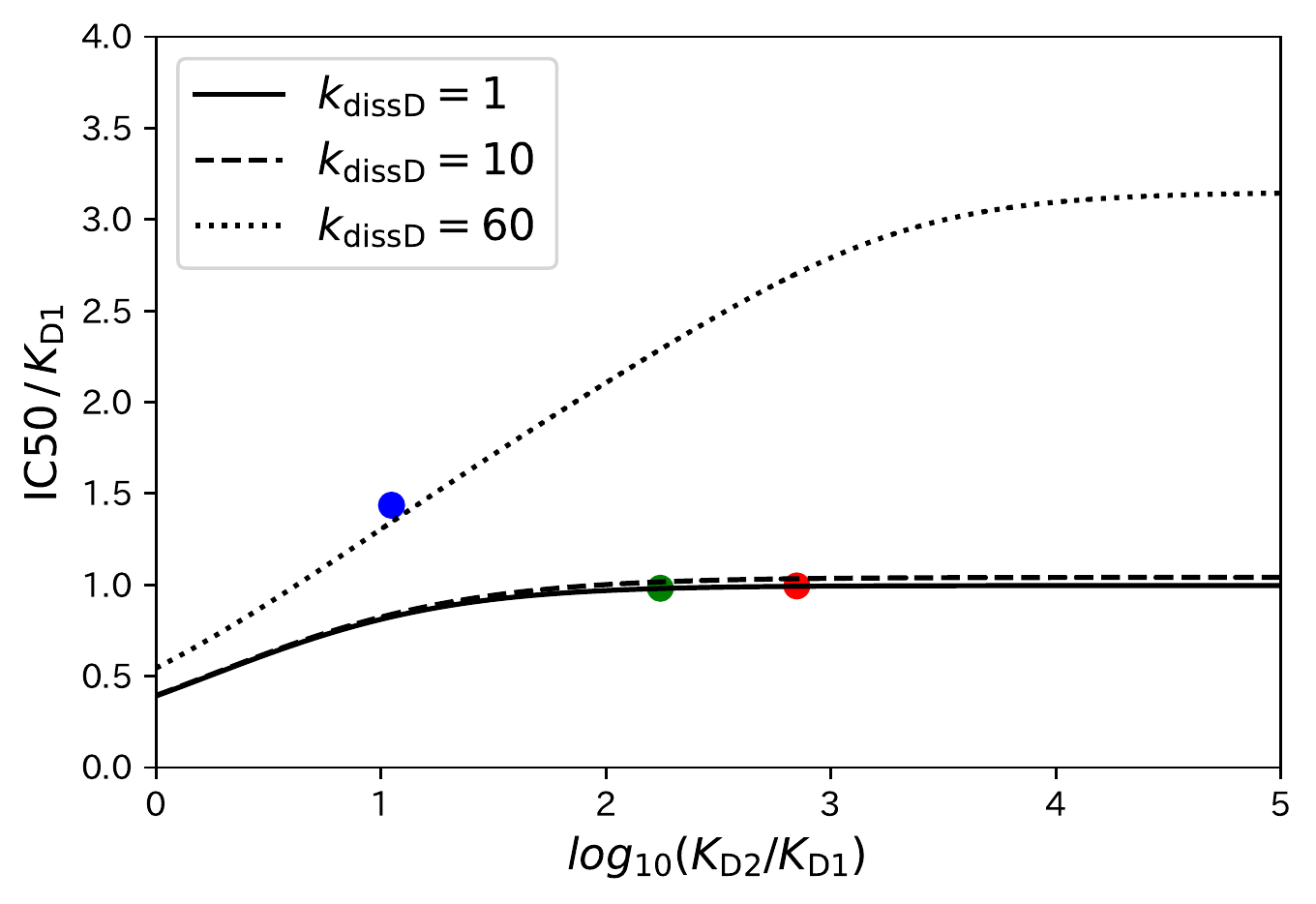}
   \subcaption{$\textrm{IC}_{50}$}  \label{fig:reciprocal_IC50}
 \end{minipage}
 \begin{minipage}[t]{0.40\linewidth}
  \includegraphics[width=0.75\linewidth]{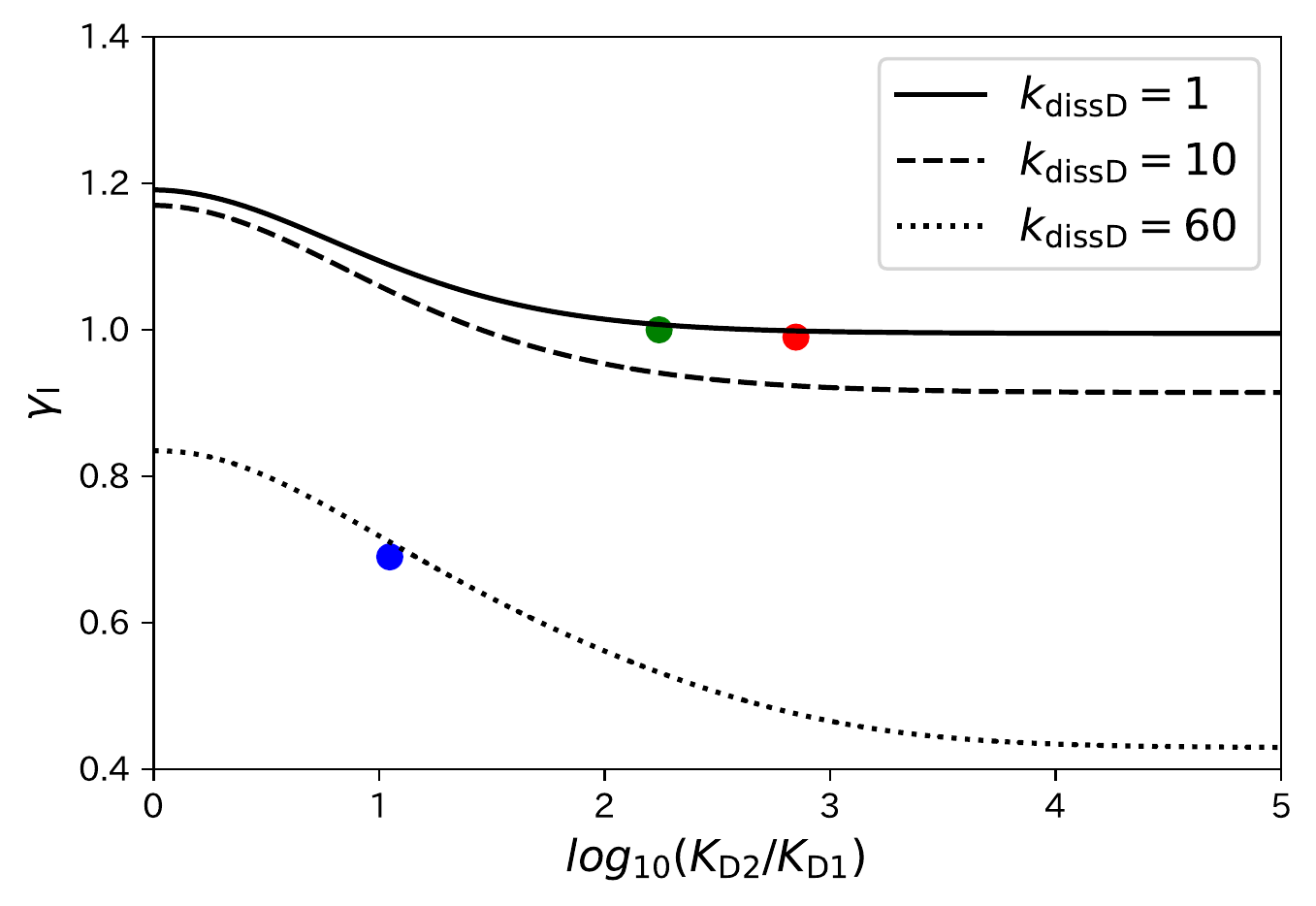}
   \subcaption{$\gamma_\mathrm{I}$} \label{fig:reciprocal_gammaI}
 \end{minipage}
 \caption{Pharmacologic parameters simulated by the competitive kinetic model with the reciprocal scheme under various settings of $\mu := K_\mathrm{D2}/K_\mathrm{D1}$. The \emph{red}, \emph{green}, and \emph{blue} points show the calculated values for cisatracurium, vecuronium, and rocuronium, respectively. }
 \label{fig:trend_reciprocal_model}    
\end{figure}

\begin{figure}[!t]
 \centering
 \begin{minipage}[t]{0.40\linewidth}
  \includegraphics[width=0.75\linewidth]{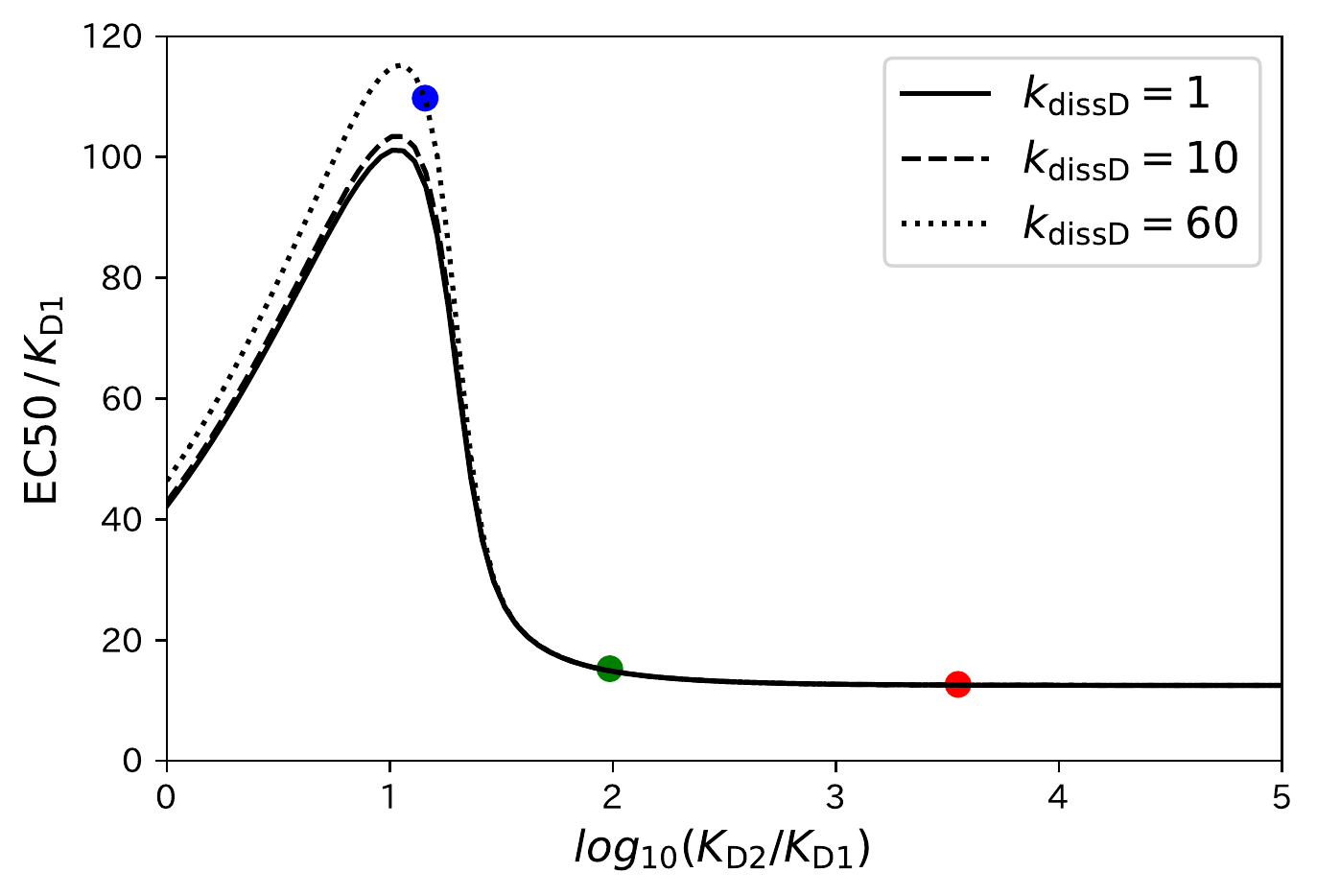}
  \subcaption{$\textrm{EC}_{50}$} \label{fig:cyclic_EC50}
  \end{minipage}
 \begin{minipage}[t]{0.40\linewidth}
  \includegraphics[width=0.75\linewidth]{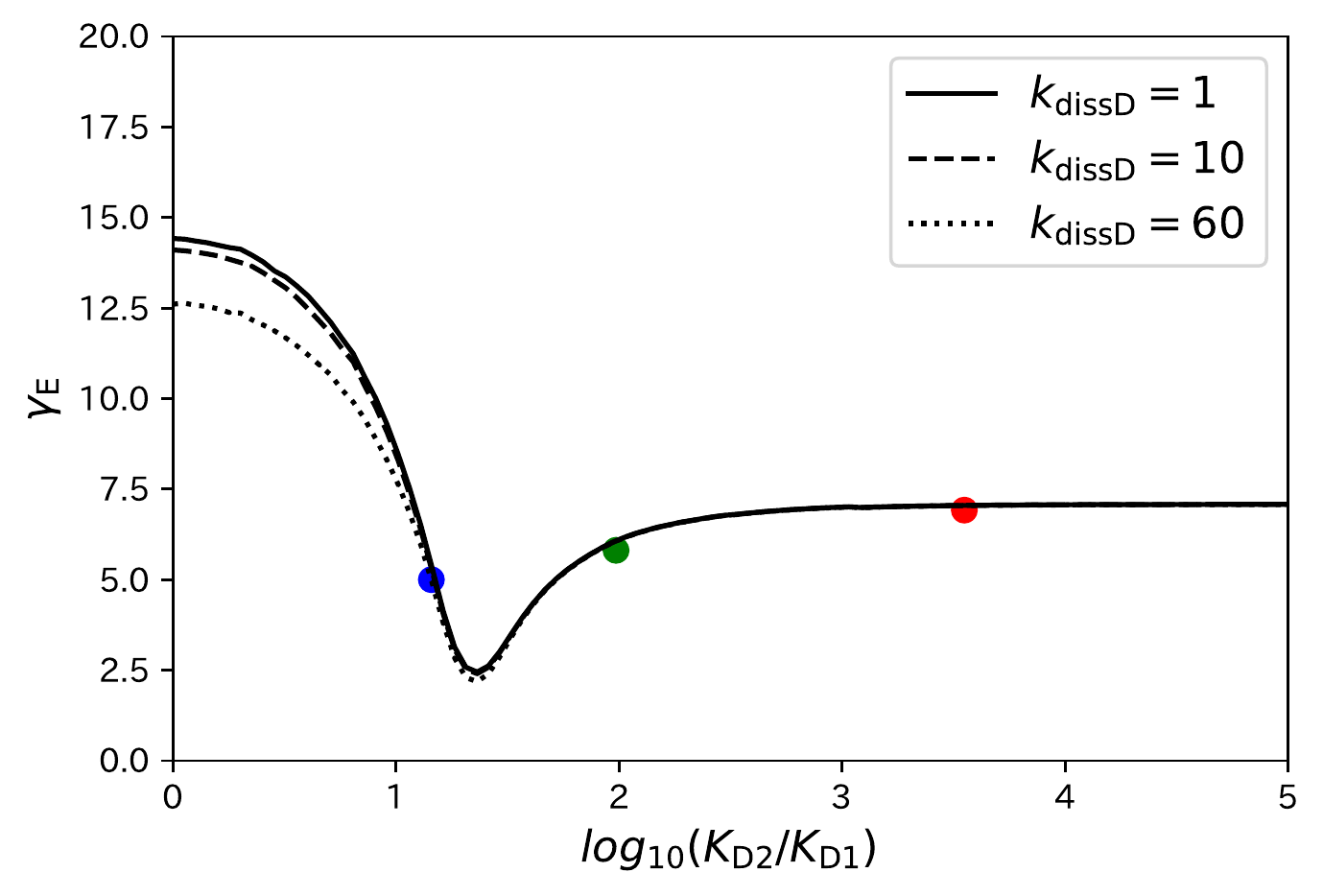}
   \subcaption{$\gamma_{E}$} \label{fig:cyclic_gammaE}
 \end{minipage}
 \begin{minipage}[t]{0.40\linewidth}
  \includegraphics[width=0.75\linewidth]{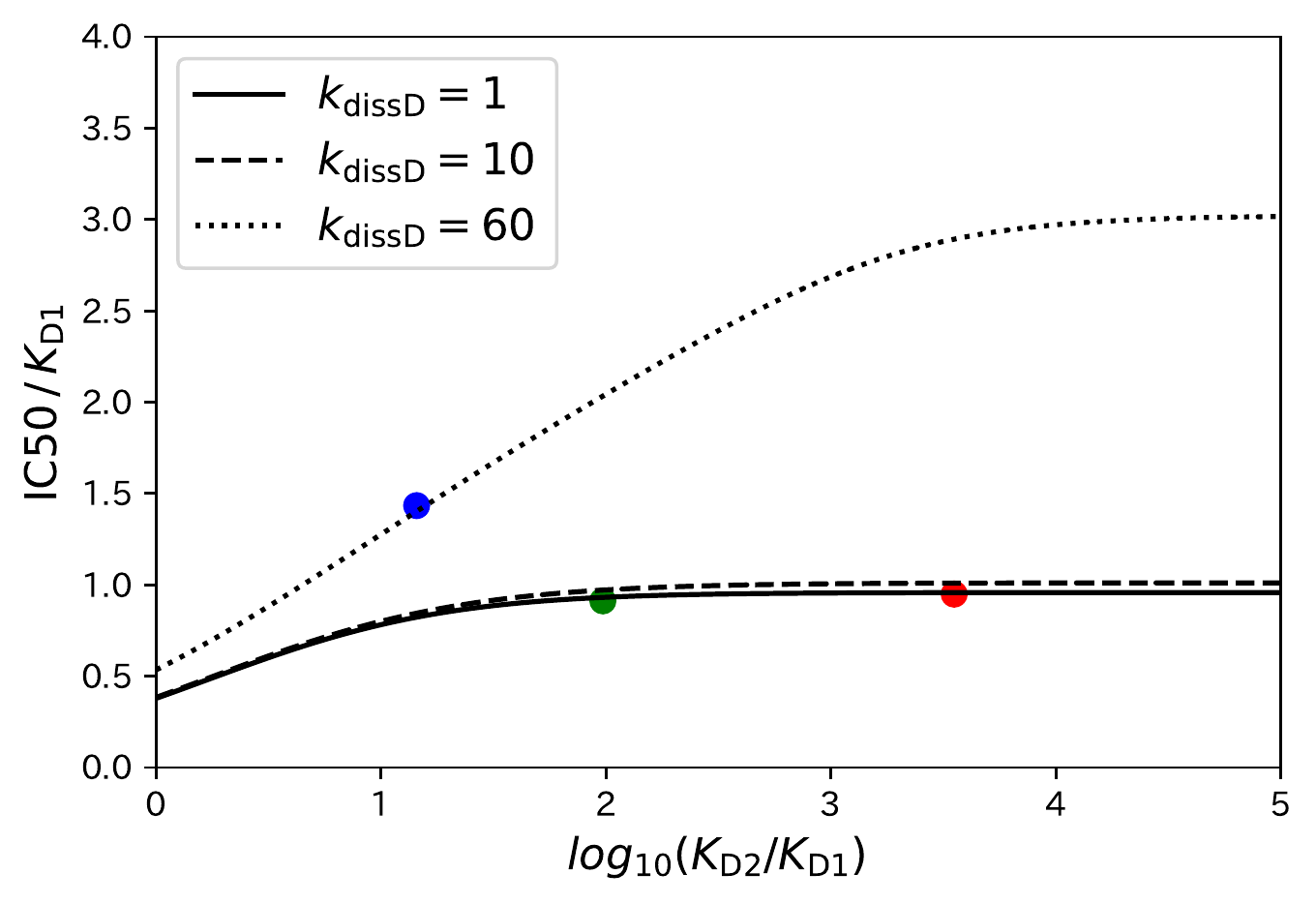}
   \subcaption{$\textrm{IC}_{50}$}  \label{fig:cyclic_IC50}
 \end{minipage}
 \begin{minipage}[t]{0.40\linewidth}
  \includegraphics[width=0.75\linewidth]{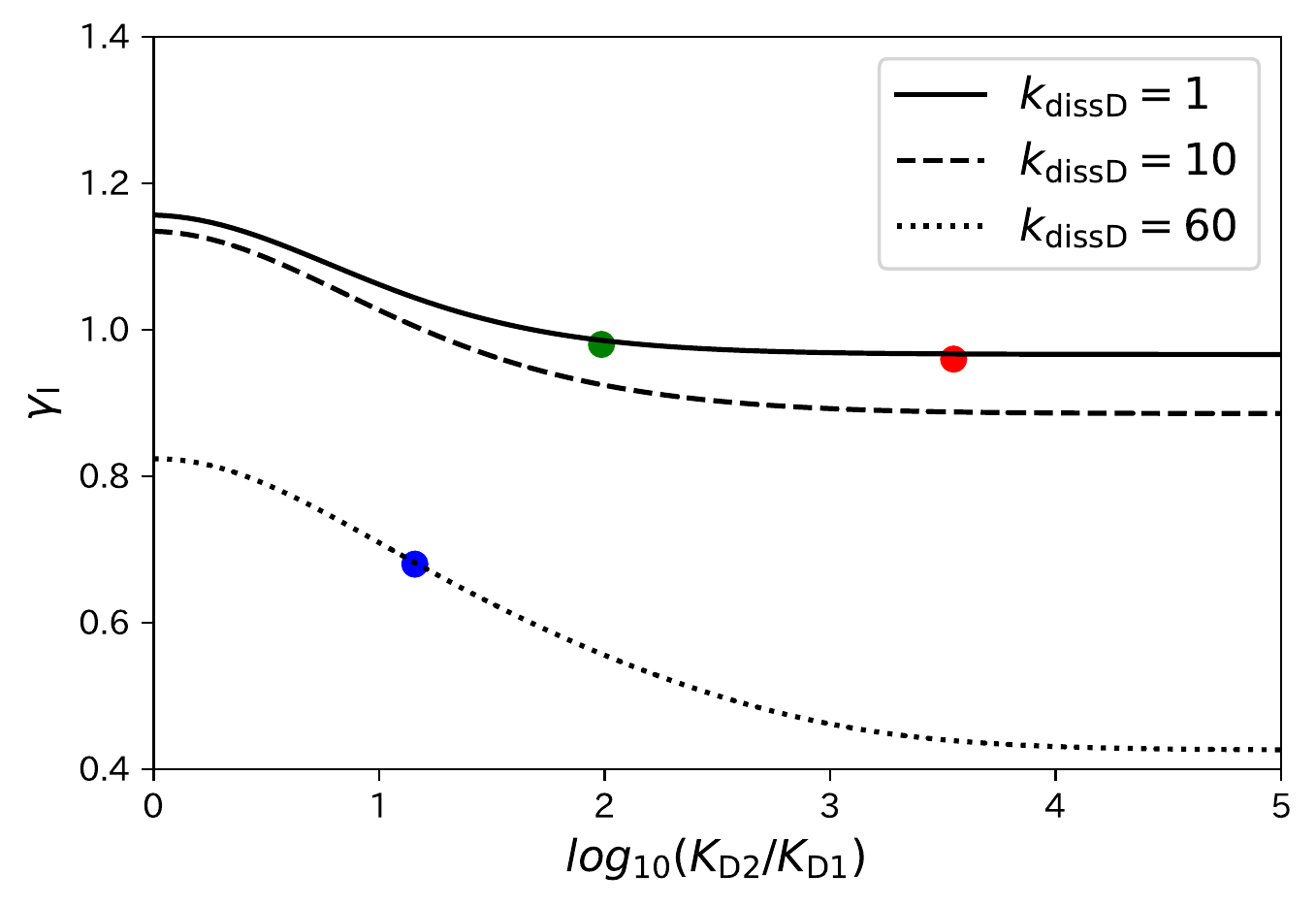}
   \subcaption{$\gamma_\mathrm{I}$} \label{fig:cyclic_gammaI}
 \end{minipage}
 \caption{Pharmacologic parameters simulated by the competitive kinetic model with cyclic scheme under various settings of $\mu := K_\mathrm{D2}/K_\mathrm{D1}$. The \emph{red}, \emph{green}, and \emph{blue} points show the calculated values for cisatracurium, vecuronium, and rocuronium, respectively.}
 \label{fig:trend_cyclic_model}    
\end{figure}

\Figrefs{fig:trend_reciprocal_model}{fig:trend_cyclic_model} show the pharmacologic parameters simulated by the competitive kinetic model with reciprocal and cyclic schemes, respectively. 
The solid, broken, and dotted lines show the simulation results with $k_\mathrm{dissD}=\SI{1.0}{s^{-1}}$, $\SI{10.0}{s^{-1}}$, and $\SI{60.0}{s^{-1}}$, respectively.
For \emph{in vitro} results, it can be seen that simulation results highly depend on  the value of  $k_\mathrm{dissD}$ and  
 the value of $\gamma_\mathrm{I}$ decreases as the increase of $k_\mathrm{dissD}$. 
This is due to the dissociation of  NDNB molecules from AChRs, which has been experimentally observed in \cite{demazumder01}, and is important to explain the low $\gamma_\mathrm{I}$ for rocuronium, which cannot be described by the two-site binding model. 
Furthermore, regarding \emph{in vivo} effects, it can be seen that the values of $\mathrm{EC}_{50}/K_\mathrm{D1}$ take the maximum values near $\mu = 10$, whereas it monotonically increases in the case of the two-site binding model. 
As a result, in contrast to the results of the two-site binding model, the site-selectivity $\mu$ for rocuronium takes the lowest values ($\mu=11.1$ for reciprocal scheme and  $\mu=14.4$ for cyclic scheme) among the three considered NDNBs. 
This is consistent with the finding for mouse AChRs in \cite{liu09} that the site-selectivity for rocuronium is lowest among the three NDNBs. 
Finally, when comparing the reciprocal and cyclic gating schemes, it can be seen that results for the cyclic scheme (\figrefs{fig:cyclic_EC50}{fig:cyclic_gammaE}) are less dependent on the value of $k_\mathrm{dissD}$ than for the reciprocal scheme. 
Its implication will be further discussed in \secref{sec:discussion}.

\section{Discussion}
\label{sec:discussion}

%

%
Regarding the difference between \emph{in vivo} and \emph{in vitro}  effects of NDNBs, it is well known that neuromuscular transmission has a high margin of safety \cite{paton67} due to copious density of AChRs, that is,  only a small fraction of AChRs need to be activated to cause muscle contraction, and more than $\SI{80}{\%}$ of the AChRs must be occupied by NDNBs before any diminution can be seen in twitch strength. 
In our simulations, neuromuscular response is calculated in the two step shown in \figref{fig:model_framework}, i.e., 1) calculation of the fraction of activated AChRs and 2) calculation of twitch strength induced by the activation of AChRs. 
In this simulation, the second step is responsible for explaining the margin of safety, and the parameters $\ce{[R^\ast]_{50}}$ and $\gamma_\mathrm{A}$ in \eref{eq:twitch_strength} determine the amount of activated AChRs needed to induce muscle contraction. 
Here,  although the low potency of rocuronium could be understood as if the margin of safety was higher for rocuronium than for cisatracurium and vecuronium, the properties of an NDNB  would not affect the second step mentioned above. 
Thus, in this paper, we aim to discuss how the apparent difference in the margin of safety can be explained through the difference in the first step of the above framework.

With this aim, this paper compared the three different model structures for the first step in \figref{fig:model_framework}: a) two-site binding model,  b) competitive kinetic model with reciprocal gating scheme, and c) competitive kinetic model with cyclic gating scheme. 
Among them, the two-site binding model is the most basic one, and it has been used to describe concentration-effect relationships obtained by \emph{in vitro} experiments \citep{sine81,colquhoun88,liu08,liu09}. 
{\color{red}
When this model is used, it is assumed that the extent of depression in the fraction of activated AChRs is proportional to 
the receptor occupancy of AChRs by NDNB. 
However, this assumption is not valid for \emph{in vivo} environment and there is a nonlinear relationship \cite{hoshino21}. 
Furthermore, this relationship depends on the property of the NDNBs.
To represent this nonlinearity in a proper way, this paper used the competitive kinetic model, which represents the dynamics of activation of AChRs by ACh molecules. 
However, we found that using reciprocal scheme results in an unrealistically slow time course of activation of AChRs and thus seems not appropriate. 
In contrast, the cyclic scheme resolves this issue and is preferred.  
}

{\color{red}
While we assumed that $k_\mathrm{dissD}=k_{\mathrm{dissD}1}=k_{\mathrm{dissD}2}$, this is not an a priori assumption validated based on literature or prior knowledge.
However, relaxing this assumption and separately optimizing the parameters $k_{\mathrm{dissD}1}$ and $k_{\mathrm{dissD}2}$ did not improve the results in \tabref{tab:results} and \figref{fig:time_course}.  
Thus, the above assumption does not affect our main conclusion that the cyclic scheme is more appropriate than the reciprocal model in describing \emph{in vivo} and \emph{in vitro} experimental results.  
It should be noted that the results here do not necessarily indicate that $k_\mathrm{dissD1}$ and $k_\mathrm{dissD2}$ are indeed equal. 
Rather, from the fact that these parameters do not largely affect the optimization results, it is implied that the values of $k_\mathrm{dissD1}$ and $k_\mathrm{dissD2}$ cannot be determined separately with sufficient reliability by the optimization procedure considered in this paper. 
}

%
%
%
{\color{red}
Regarding dynamic effect of the activation of AChRs, in} our previous study \citep{hoshino21}, we theoretically and numerically analyzed the relationship between the fraction of activated AChRs and the receptor occupancy by NDNB molecules.
As a result, it has been shown that the fraction of activated AChRs simulated by the competitive kinetic model get closer to that described by the two-site binding model as 1) the concentration of ACh becomes higher and 2) the dissociation rate constant $k_\mathrm{diss}$ becomes smaller.
Conversely,  small ACh concentration or large rate constant $k_\mathrm{diss}$ is necessary for explaining \emph{
in vivo} experimental results. 
Furthermore, it has been found in \cite{hoshino21} that the difference between \emph{in vivo}  and \emph{in vitro} simulations becomes more prominent as the value of $\mu=K_\mathrm{D2}/K_\mathrm{D1}$ decreases, i.e., as the site-selectivity becomes small. 
Thus, as shown in \figrefs{fig:trend_reciprocal_model}{fig:trend_cyclic_model},  the value of $\mathrm{EC}_{50}/K_\mathrm{D1}$ takes a maximum value because it increases with the decrease in $\mu$ in the range of $\log_{10}\mu > 1$ due to the above reason and decreases with the decrease in $\mu$ in the range of $0<\log_{10}\mu<1$  due to a change in the receptor occupancy as described by the two-site binding model (as shown in \figref{fig:trend_two-site_model}).
Owing to the fact that the value of $\mathrm{EC}_{50}/K_\mathrm{D1}$ takes a large value near  $\mu=10$, where the value of  $\mathrm{IC}_{50}/K_\mathrm{D1}$ is relatively low as shown in \figrefs{fig:trend_reciprocal_model}{fig:trend_cyclic_model}, it is possible to explain the high ratio of $\mathrm{EC}_{50}/\mathrm{IC}_{50}$  for rocuronium. 
{\color{red}
%
%
%
However,}  it can be seen that  the reason for the low potency of rocuronium is differently explained by the models with reciprocal and cyclic gating schemes. 
When the reciprocal scheme is used,  the  \emph{in vivo} effects ($\mathrm{EC}_{50}$ and $\gamma_\mathrm{E}$ in  \figref{fig:trend_reciprocal_model})  highly depend on the dissociation rate constant $k_\mathrm{dissD}$, 
and thus,  large $k_\mathrm{dissD}$ is necessary to explain  high $\mathrm{EC}_{50}$  for rocuronium. 
On the other hand, when the cyclic scheme is used (\figref{fig:trend_cyclic_model}),  the \emph{in vivo} results are less dependent on  the value of $k_\mathrm{dissD}$. 
Thus, if the cyclic scheme is appropriate, it follows that  $k_\mathrm{dissD}$ is not an important factor to explain the low potency of rocuronium, while  it is still important to explain the low $\gamma_\mathrm{I}$ for rocuronium. 
Since it  has been found in \cite{hoshino21} that either small ACh concentration or large  $k_\mathrm{dissD}$ is necessary to  explain  the difference between the simulation results of the two-site binding model and the competitive kinetic model,  small ACh  concentration would be a key factor needed to explain the low potency of rocuronium.

Although the cyclic gating scheme is preferred in this paper, the reciprocal gating scheme has been a widely accepted model \cite{akk96,grosman01,auerbach12,gupta17}. 
A key fact that has been supported the reciprocal gating scheme is that the affinity of the AChR for ACh is much higher in the open than in the closed state \cite{auerbach12,gupta17}, which can lead to the thought that ACh would not dissociation from AChRs while the channel is open. 
Interestingly, however, with the estimated parameters of the cyclic model, the affinity in the open state ($1/K_\mathrm{A}^\ast= \SI{5.43e7}{M^{-1}}$) is higher than that in the closed state ($1/K_\mathrm{A}= \SI{2.25e6}{M^{-1}}$ ) and are consistent with the above fact. 
That is, the high affinity at the open state does not exclude the possibility of a cyclic model. 
However, there is a discrepancy between the dissociation rate $k_\mathrm{dissA}^\ast$ estimated in this paper ({\color{red}$\SI{1.70e4}{s^{-1}}$}) and reported in \cite{grosman01} ($\SI{24}{s^{-1}}$). 
Since the constant  reported in \cite{grosman01} is estimated based on the premise of the reciprocal model,  it may be possible to reconcile the discrepancy by re-estimating the constant using the cyclic scheme. 
However, it is beyond the scope of this paper and is in the future work.
Thus, further consideration is needed to discuss whether the cyclic gating scheme is appropriate or not.

Finally, it should be noted that the method of parameter estimation and the estimated results are dependent on the following simplifying assumptions. 
1) In the kinetic simulation performed in this paper, we ignored the effect of the three-dimensional structure of the synapse as modeled in \cite{bartol91,stiles96} and simply assumed that both ACh and AChRs are distributed uniformly in the synaptic cleft.
2) The extent of plasma protein binding, which is different for each NDNB \citep{roy04}, was not considered to simulate the \emph{in vivo} effects. 
Although only the unbound NDNB molecules can cross cell membranes and reach the postsynaptic membrane at the neuromuscular junction, we simply considered the effect-site concentration in the PKPD modeling as the NDNB concentration at the neuromucsular junction.
3) All the kinetic constants were assumed to be the same between \emph{in vivo} and \emph{in vitro} environments. For example, it has been reported that the kinetic constants for NDNBs at physiological temperatures around $\SI{37}{\degreeCelsius}$ are different from those at room temperatures around $\SI{25}{\degreeCelsius}$ \citep{demazumder08}.  
4) Several parameters are directly taken from literature without any correction. For example, the concentration $\ce{[R]}_\mathrm{total}$ of AChRs is based on the number of AChRs at the end plates of human deltoid muscle \citep{fambrough73} and the volume of the synaptic cleft of rat diaphragm \citep{rosenberry79}, and thus it may be different from the value for human adductor pollicis muscle for which \emph{in vivo} experimental results have been obtained. 
Regardless of these assumptions, the obtained model and simulation results of this paper are useful for exploring the molecular mechanisms of the relationship between \emph{in vivo} and \emph{in vitro} effects of NDNBs.

\section{Conclusions} \label{sec:conclusion}

This paper addressed simultaneous modeling of \emph{in vivo} and \emph{in vitro} effects of NDNBs.
In particular, we explored a suitable model structure and its parameters to reconcile the fact that rocuronium is less potent at inducing muscle relaxation \emph{in vivo} than directly predicted from \emph{in vitro} experiments. 
By comparing the results of parameter estimation for three candidate models, it was shown that the competitive kinetic model with the cyclic gating scheme best described both the \emph{in vivo} and \emph{in vitro} experimental data. 
It was found that the above apparent discrepancy can be resolved  if we assume that the \emph{in vivo} concentration of ACh is relatively low to activate only a part of AChRs, whereas more than $\SI{95}{\%}$ of AChRs are activated during \emph{in vitro} experiments, and that the site-selectivity is smaller for rocuronium than those for cisatracurium and vecuronium. 

{\color{red}
\section*{Appendix A: Model equation of the cyclic scheme} 

The ordinary differential equation model for the cyclic scheme is given as follows:
\begin{subequations} \label{eq:na-model}
\begin{align}
& \dfrac{\dd}{\dd t}\ce{[A]} =  - {\color{red}  k_\mathrm{decay} } \ce{[A]}
  + (k_\mathrm{dissA1}^\ast+k_\mathrm{dissA2}^\ast)\ce{[ARA^\ast]}  
   - k_\mathrm{assocA1}^\ast \ce{[ORA^\ast]} - k_\mathrm{assocA2}^\ast \ce{[ARO^\ast]}  \notag \\
  &\hspace{13mm} + k_\mathrm{dissA1} (\ce{[ARA]} + \ce{[ARD]} + \ce{[ARO]}) 
   - k_\mathrm{assocA1} \ce{[A]} (\ce{[ORA]}	 + \ce{[ORD]} + \ce{[ORO]})  \notag \\
  &\hspace{13mm} + k_\mathrm{dissA2} ( \ce{[ARA]} + \ce{[DRA]} + \ce{[ORA]}) 
   - k_\mathrm{assocA2} \ce{[A]} (\ce{[ARO]}	 + \ce{[DRO]} + \ce{[ORO]}),  \\ 
 &  \dfrac{\dd}{\dd t}\ce{[ARA^\ast]} =  k_\mathrm{open} \ce{[ARA]} - (k_\mathrm{dissA1}^\ast+k_\mathrm{dissA2}^\ast)\ce{[ARA^\ast]} 
  + k_\mathrm{assocA1}^\ast \ce{[ORA^\ast]} + k_\mathrm{assocA2}^\ast \ce{[ARO^\ast]} \notag \\
  & \hspace{19mm} - k_\mathrm{d+} \ce{[ARA^\ast]}  + k_\mathrm{d-} \ce{[RD]},  
\end{align}
\begin{align}
 &   \dfrac{\dd}{\dd t}\ce{[ARO^\ast]} =  - k_\mathrm{close}\ \ce{[ARO^\ast]} + k_\mathrm{dissA1}^\ast \ce{[ARA^\ast]}  - k_\mathrm{assocA1}^\ast \ce{[ORA^\ast]},   \\
 &  \dfrac{\dd}{\dd t}\ce{[ORA^\ast]} =  - k_\mathrm{close}\ \ce{[ORA^\ast]}  + k_\mathrm{dissA2}^\ast \ce{[ARA^\ast]}  - k_\mathrm{assocA2}^\ast \ce{[ARO^\ast]} , \\
& \dfrac{\dd}{\dd t}\ce{[ARA]} = 
   - k_\mathrm{open} \ce{[ARA]}
   k_\mathrm{assocA1} \ce{[ORA][A]} -  k_\mathrm{dissA1}\ce{[ARA]} 
   + k_\mathrm{assocA2} \ce{[ARO][A]} -  k_\mathrm{dissA2}\ce{[ARA]}  \\
& \dfrac{\dd}{\dd t}\ce{[DRD]} = 
 k_\mathrm{assocD1} \ce{[ORD][D]} -  k_\mathrm{dissD1}\ce{[DRD]} 
  +  k_\mathrm{assocD2} \ce{[DRO][D]} -  k_\mathrm{dissD2}\ce{[DRD]}, \\ 
& \dfrac{\dd}{\dd t}\ce{[ARD]} =  
 k_\mathrm{assocA1} \ce{[ORD][A]} -  k_\mathrm{dissA1}\ce{[ARD]} 
  + k_\mathrm{assocD2} \ce{[ARO][D]} -  k_\mathrm{dissD2}\ce{[ARD]}, \\ 
& \dfrac{\dd}{\dd t}\ce{[DRA]} = 
  k_\mathrm{assocD1} \ce{[ORA][D]} -  k_\mathrm{dissD1}\ce{[DRA]} 
  +  k_\mathrm{assocA2} \ce{[DRO][A]} -  k_\mathrm{dissA2}\ce{[DRA]}, 
\\
 & \dfrac{\dd}{\dd t}\ce{[ARO]} =  k_\mathrm{assocA1} \ce{[ORO][A]} - k_\mathrm{dissA1}\ce{[ARO]} 
  + k_\mathrm{dissA2} \ce{[ARA]} - k_\mathrm{assocA2} \ce{[ARO][A]} \notag \\ 
  &\hspace{17mm} + k_\mathrm{dissD2} \ce{[ARD]} - k_\mathrm{assocD2} \ce{[ARO] [D]}, \\
& \dfrac{\dd}{\dd t}\ce{[ORA]} =  k_\mathrm{assocA2} \ce{[ORO][A]} - k_\mathrm{dissA2}\ce{[ORA]}
  + k_\mathrm{dissA1} \ce{[ARA]} - k_\mathrm{assocA1} \ce{[ORA][A]} \notag \\ 
 &\hspace{17mm}  + k_\mathrm{dissD1} \ce{[DRA]} - k_\mathrm{assocD1} \ce{[ORA] [D]}, \\
& \dfrac{\dd}{\dd t}\ce{[DRO]} =  k_\mathrm{assocD1} \ce{[ORO][D]} - k_\mathrm{dissD1}\ce{[DRO]} 
  + k_\mathrm{dissD2} \ce{[DRD]} - k_\mathrm{assocD2} \ce{[DRO] [D]} 
 \notag \\
  &\hspace{17mm} + k_\mathrm{dissA2} \ce{[DRA]} - k_\mathrm{assocA2} \ce{[DRO][A]}, 
  \\ 
& \dfrac{\dd}{\dd t}\ce{[ORD]} = k_\mathrm{assocD2} \ce{[ORO][D]} - k_\mathrm{dissD2}\ce{[ORD]} 
+ k_\mathrm{dissD1} \ce{[DRD]} - k_\mathrm{assocD1} \ce{[ORD] [D]}   \notag \\
  &\hspace{17mm} + k_\mathrm{dissA1} \ce{[ARD]} - k_\mathrm{assocA1} \ce{[ORD][A]},  \\
& \dfrac{\dd}{\dd t}\ce{[RD]} = k_\mathrm{d+} \ce{[ARA^\ast]}  - k_\mathrm{d-} \ce{[RD]}
\end{align}
\end{subequations}
where the symbol $\ce{[X]}$ stands for the concentration of a substance $\ce{X}$. 
The concentration $\ce{[ORO]}$ of the unoccupied AChR is given by 
\begin{align}
 \ce{[ORO]} =& \ce{[R]}_\mathrm{total}  {\color{red}- \ce{[ARA^\ast]} } - \ce{[ARA]} - \ce{[DRD]} - \ce{[ARD]} - \ce{[DRA]} 
  -  \ce{[ARO]} - \ce{[ORA]} - \ce{[DRO]} - \ce{[ORD]},
\end{align}
where $\ce{[R]}_\mathrm{total}$ stands for the concentration of the post-junctional AChRs in the synaptic cleft. 
The initial values of the states can be given by
\begin{align}
& \ce{[A]}|_{t=0} = \ce{[A]}_\mathrm{init},   \quad  \ce{[ARA^\ast]}|_{t=0} = \ce{[ARO^\ast]}|_{t=0}  =  \ce{[ORA^\ast]}|_{t=0}  =  0, \\ 
&  \ce{[ARA]}|_{t=0} = \ce{[ARO]}|_{t=0}  =  \ce{[ORA]}|_{t=0}  =  \ce{[ARD]}|_{t=0} = \ce{[DRA]}|_{t=0}  =  \ce{[RD]}|_{t=0} =  0, \\
&  \ce{[DRD]}|_{t=0} = \dfrac{ \ce{[D]}^2 K_\mathrm{A1} K_\mathrm{A2} }{ \Delta }, \quad 
    \ce{[DRO]}|_{t=0} = \dfrac{ \ce{[D]} K_\mathrm{D2} K_\mathrm{A1} K_\mathrm{A2} }{ \Delta }, \quad 
    \ce{[ORD]}|_{t=0} = \dfrac{ \ce{[D]} K_\mathrm{D1} K_\mathrm{A1} K_\mathrm{A2} }{ \Delta }. 
\end{align}
with $\Delta := ( \ce{[D]} K_\mathrm{A1} + K_\mathrm{D1} K_\mathrm{A1})( \ce{[D]} K_\mathrm{A2} + K_\mathrm{D2} K_\mathrm{A2}) $. 
}

\section*{Acknowledgements}

 This work was partially supported by Grant-in-Aid for Scientific Research (KAKENHI) from the Japan Society for Promotion of Science (\#20K04553).

\section*{Competing Interest}

The authors declare that they have no competing interests.

\section*{Ethics declaration}

This study was conducted by theoretical investigations and computer-based simulations. 
As such, the data employed in this study did not require ethical approval.


\bibliographystyle{elsarticle-num}
\bibliography{hoshino_anesthesia}   

\end{document}